\documentclass[aps,prd,preprintnumbers,showpacs,showkeys,nofootinbib,
superscriptaddress,fleqn,floatfix,tightenlines,10pt]{revtex4-1}
\usepackage{amsmath,amsfonts,amssymb,amscd,amsxtra,amsthm}
\usepackage{graphicx}  
\usepackage{epstopdf}
\usepackage{dcolumn}  
\usepackage{bm}          
\usepackage{slashed}
\usepackage{cancel}
\usepackage{float}
\usepackage{mathtools}
\usepackage{amsbsy}
\usepackage{amstext}

\usepackage[utf8]{inputenc} 
\usepackage{booktabs}
\usepackage[normalem]{ulem} 
\usepackage[dvipsnames]{xcolor} 
\usepackage{tabularx}
\usepackage{enumitem}  
\usepackage{array} 
\usepackage{slashed}
\usepackage{tikz}
\usepackage{float}
\usepackage{multirow}
\renewcommand\sout{\bgroup \color{red} \ULdepth=-.5ex \ULset}

\makeatletter

\begin{document}  
 \preprint{INHA-NTG-07/2021}
\title{Structure of the $\Omega$ baryon and the kaon cloud}
\author{Yu-Son Jun}
\email[E-mail: ]{ysjun@inha.edu}
\affiliation{Department of Physics, Inha University, Incheon 22212,
Republic of Korea}

\author{Hyun-Chul Kim}
\email[E-mail: ]{hchkim@inha.ac.kr}
\affiliation{Department of Physics, Inha University, Incheon 22212,
Republic of Korea}
\affiliation{School of Physics, Korea Institute for Advanced Study
(KIAS), Seoul 02455, Republic of Korea}

\author{June-Young Kim}
\email[E-mail: ]{Jun-Young.Kim@ruhr-uni-bochum.de}
\affiliation{Institut f\"ur Theoretische Physik II, Ruhr-Universit\"at
  Bochum, D-44780 Bochum, Germany}

\author{Jung-Min Suh}
\email[E-mail: ]{suhjungmin@inha.edu}
\affiliation{Department of Physics, Inha University, Incheon 22212,
Republic of Korea}

\date{\today}
\begin{abstract}
We investigate the effects of the kaon cloud on the electromagnetic
and axial-vector form factors of the $\Omega^-$ baryon within the 
framework of the chiral quark-soliton model. We first derive the
profile function of the chiral soliton in such a way that the kaon
Yukawa tail is properly produced self-consistently. Then, we compute
the electromagnetic form factors of the $\Omega^-$ baryon. The
results for the electromagnetic form factors are compared with the
lattice data. We find that the results with the kaon tail employed are
in better agreement with the lattice data, compared with those
obtained with the pion tail. We also study the axial-vector form
factors of the $\Omega^-$ baryon, examining the effects of the kaon
cloud.  
\end{abstract}
\pacs{}
 \keywords{}
\maketitle
\section{Introduction}
The lowest-lying $\Omega$ baryon is the strangest particle among all 
hadrons. It is also a member of the baryon decuplet, which was found
the latest~\cite{Barnes:1964pd} among the lowest-lying decuplet
baryons. The excited state $\Omega(2250)^-$ 
was for the first time found after two decades
passed~\cite{Biagi:1985rn, Aston:1987bb, Aston:1988yn}. More 
interestingly, the first excited state $\Omega(2012)^-$ was found only
very recently~\cite{Yelton:2018mag, Jia:2019eav}. The Particle Data
Group (PDG) summarizes the existence of the four excited
states~\cite{Zyla:2020zbs} only. This indicates that the $\Omega$  
baryons have been least studied among hyperons. A very recent
measurement of the correlation function for the proton ($p$)-$\Omega$
interaction has triggered interest in these $\Omega$
baryons~\cite{Acharya:2020asf}. It reveals that the $p$-$\Omega$
potential is attractive even though the relative distance
between $p$ and $\Omega$ goes to zero, which implies a possible
existence of the $p$-$\Omega$ bound state~\cite{Iritani:2018sra}. In
the meanwhile, the electromagnetic (EM) form factors and axial charges
of the ground-state $\Omega$ baryon were computed in lattice quantum  
chromodynamics (QCD)~\cite{Alexandrou:2010jv, Alexandrou:2016xok}. 

The nucleon has been known to be dressed by the pion
cloud in various contexts~\cite{Yennie:1957, Frazer:1959gy, 
  Frazer:1960zza, Frazer:1960zzb, CohenTannoudji:1972gd,
  Anselm:1972ir, Adkins:1983hy, Frankfurt:1988nt, Hammer:2006mw,
  Meissner:2007tp, Jenkovszky:2017efs}. The pion cloud can be
qualitatively understood as the $q\bar{q}$ excitations within the
framework of QCD~\cite{Frankfurt:1988nt} or as the pion loop in
effective field theory~\cite{Meissner:2007tp}. The large $N_c$ (the
number of colors) limit of QCD provides yet a lucid point of view on
the pion cloud of the nucleon~\cite{Witten:1979kh}. The nucleon emerges
as states of the $N_c$ valence quarks bound by the pion mean field,
which is produced by the presence of the $N_c$ valence quarks. This
pion mean field can be regarded as the pion cloud. In the Skyrme model
with the finite pion mass included, which is the simplest case of the
pion mean-field approach, the expectation value of the pion field 
in a nucleon shows an asymptotic behavior: $\langle
\pi^a (r)\rangle \sim  e^{-m_\pi r}x^i r^{-2}\langle \sigma_i
\tau^a\rangle$~\cite{Adkins:1983hy}. A more realistic pion-mean field
approach such as the chiral quark-soliton
model($\chi$QSM)~\cite{Diakonov:1987ty,Diakonov:1997sj} provides also  
the pion cloud from the Dirac-sea polarization or the Dirac-sea 
continuum~\cite{Christov:1995vm, Kim:1995mr}. 

On the other hand, the $\Omega$ baryon is not coupled to the pion,
since it is an isosinglet baryon and consists only of the strange
valence quarks. However, the SU(3) baryons are constructed in the
$\chi$QSM by embedding the SU(2) soliton into flavor SU(3) space, so 
all the lowest-lying baryons in the model have the pion
tail~\cite{Witten:1983tw, Jain:1984gp, Diakonov:1997sj} in
common. While this is theoretically consistent, it does not reflect
phenomenology properly. Since the hyperons contain the strange quark,
the kaon cloud should naturally come into play in describing the
structure of them. Even in the case of the nucleon, the kaon cloud 
provides certain effects when the strangeness in the nucleon is
examined~\cite{Watabe:1995ts, Kim:1996vj, 
  Silva:2001st, Silva:2002ej, Silva:2005qm}. However, in any chiral
solitonic approaches, it is impossible to consider both the pion and
kaon clouds in a compatible manner. This means that one should choose
either the pion cloud or the kaon one, depending on which observables
of the hyperons are investigated. For example, the electric form
factors of the neutron or neutral hyperons are rather sensitive to the
meson tails. Fortunately, the $\Omega$ baryon contains three
strange valence quarks and it is not coupled to the pions, so we can 
choose the kaon tail in place of the pion one. Of course, one may
still raise a question about the contributions from the $\eta$ and
$\eta'$ clouds, since these two pseudoscalar mesons also contain
hidden strangeness, so that they can be coupled to
the $\Omega$ baryon. The coupling between $\eta$ and $\Omega$ is known
to be rather small~\cite{Yang:2018idi}, so the effect of the $\eta$ is
expected to be also small on the $\Omega$ baryon. Since the mass of
$\eta'$ is much larger than the pseudoscalar meson octet, its effect
should be suppressed. Thus, as far as we are only concerned with the
$\Omega$ baryon, the present approach will still shed light on
understanding the effects of kaon cloud or the kaon tail on the
structure of the $\Omega$ baryon.  

In the present work, we will investigate the EM and axial-vector
structures of the $\Omega^-$ baryon with the effects of the kaon cloud  
considered. We first review the kaon properties based on the effective
chiral action and fix the relevant parameters such as the pion decay
constant, pion and kaon masses.  We will then introduce the effective
mass of the current strange quark in the Dirac Hamiltonian and solve
the classical equation of motion, so that we are able to derive the
profile function of the chiral soliton with the proper kaon tail. Then
we will perform the symmetry-conserving zero-mode
quantization~\cite{Praszalowicz:1998jm}. We will focus on examining
the EM and axial-vector form factors of the $\Omega$ baryon, taking
into account the $1/N_c$ rotational corrections on them. Since the
effects of the explicit breaking of flavor SU(3) symmetry are
rather small, we do not include them in the present work. Finally, we
will compare the present results with those from lattice QCD and
explicitly show that the kaon cloud plays an essential role in
describing the $\Omega$ baryon.  

The present paper is organized as follows: In Section II, we start
from the effective chiral action to study the properties of the
kaon. Expanding the effective action by using the background field
method, one can find the kaon propagator of which the pole yields the
mass of the kaon and its residue becomes its decay constant. This
process is similar to a usual renormalization 
of a particle. Then we proceed to the solitonic sector and derive the
equation of motion of which the solution leads to the meson mean
field. The kaon cloud can be incorporated by using the corresponding
quark mass inside the one-body Dirac Hamiltonian. The classical
solution, which is identified as the classical mass of the $\Omega$
baryon, possesses the Yukawa tail with the proper kaon mass. This will
play a key role in describing the EM and axial-vector form factors of
the $\Omega$. In Section III, we briefly recapitulate the formulae
for the EM and axial-vector form factors. In Section IV, we present
the numerical results and discuss them, comparing them with those from
lattice QCD. The last Section is devoted to the summary and conclusion
of the present work. In Appendices, we compile the expressions for the
EM and axial-vector form factors, derived from the $\chi$QSM.  

\section{Kaon properties and kaon cloud}
The effective chiral action and its partition function in Euclidean
space are given as  
\begin{align}
&Z_{\mathrm{eff}} = \int \mathcal{D}\psi^\dagger
  \mathcal{D}\psi \mathcal{D}\pi^a \mathrm{exp}\left[\int d^{4}x
  \psi^{\dagger}i D(\pi^a)\psi\right]= \int \mathcal{D} \pi^a
  \mathrm{exp}[-S_{\mathrm{eff}}], 
\label{eq:effecXac}
\end{align}
where $S_{\mathrm{eff}}$ represents the effective chiral action
expressed by 
\begin{align}
&S_{\mathrm{eff}}[\pi] = -N_{c}\mathrm{Tr\;ln}\;(i
  \slashed{\partial} +i  \hat{m}+i MU^{\gamma_{5}}(\pi^a)). 
\label{eq:effac}
\end{align}
Here, $N_{c}$ is the number of colors. $\mathrm{Tr}$ denotes the
functional trace over the space-time, spin space, and flavor
space. $M$ designates the dynamical quark mass. $\hat{m}$ is the 
matrix of the current-quark masses 
\begin{align}
\hat{m} = \left(
  \begin{array}{ccc}
m_{\mathrm{u}}    & 0 & 0 \\
0 & m_{\mathrm{d}}    & 0 \\
0 & 0 & m_{\mathrm{s}}   
  \end{array}
\right).
\end{align}
$U^{\gamma_5}$ stands for the chiral field expressed by
\begin{align}
U^{\gamma_5} = \frac{1+\gamma_5}{2} U  + \frac{1-\gamma_5}{2}
  U^\dagger  
\end{align}
with 
\begin{align}
U(x) = \exp\left[ \frac{i}{f} \lambda^a \pi^a
  (x)\right] .
\label{eq:u}
\end{align}
Here, $f$ is introduced as a scale factor that makes the argument of
Eq.~\eqref{eq:u} dimensionless. $\pi^a$ represent the
pseudo-Nambu-Goldstone fields expressed as 
\begin{align}
\lambda^a \pi^a = \left(
  \begin{array}{ccc}
\frac1{\sqrt{2}} \pi^0 + \frac1{\sqrt{6}} \eta    & \pi^+ & K^+ \\ 
\pi^- & -\frac1{\sqrt{2}} \pi^0 + \frac1{\sqrt{6}} \eta & K^0 \\
K^- & \bar{K}^0 & -\frac2{\sqrt{6}} \eta
  \end{array}
\right).
\end{align}

\subsection{Kaon properties}
We first examine the meson properties based on the effective chiral
action~\cite{Jaminon:1989wp, Jaminon:1991vz,
  Christov:1995vm}. Introducing the meson source $j^a$ 
explicitly, we write the generating functional as follows:
\begin{align}
\mathcal{Z} [j] = \int \mathcal{D} \pi^a e^{-S_{\mathrm{eff}} + j\cdot \pi},  
\end{align}
which gives the mesonic two-point correlation function as 
\begin{align}
K^{ab}(x-y) = \left. \frac{\delta^2 \ln \mathcal{Z}}{\delta j^a(x) \delta
  j^b(y)} \right|_{j=0},    
\end{align}
where $j\cdot \pi = \int d^4 x j^a (x) \pi^a (x)$. Note that in the
present model the mesons emerge as low-lying collective $\bar{\psi}
\psi$ excitations. Using the background field
method~\cite{Abbott:1981ke}, one can decompose the mesonic field
$\pi^a$ into two parts:
\begin{align}
\pi^a (x) = \pi_c^a (x) + \delta \pi^a(x),  
\end{align}
where $\pi^a$ is the solution of the classical equation of motion
\begin{align}
\frac{\delta S}{\delta \pi^a(x)} =0,
\end{align}
which is just the same as the classical value of $\pi^a$
\begin{align}
\pi_c^a (x) := \langle \pi^a\rangle =\left. \frac{\delta
  \ln\mathcal{Z}}{\delta j^a(x)}  \right|_{j=0}.
\end{align}
Then, the generating functional can be written as 
\begin{align}
\ln \mathcal{Z} =-S_{\mathrm{eff}}[\pi_c^a] + j\cdot \pi_c -\frac12
  \mathrm{Tr}\ln\left[\frac{\delta^2 S}{\delta \pi \delta \pi}\right]
  +\frac12 j\left[\frac{\delta^2 S}{\delta \pi \delta \pi}
  \right]^{-1} j , 
\label{eq:gf2}
\end{align}
where the last term is a short-handed notation given as 
\begin{align}
j\left[\frac{\delta^2 S}{\delta \pi \delta \pi}
  \right]^{-1} j = \int d^4x d^4y\,  j^a(x) \frac{\delta^2 S}{\delta
  \pi^a(x) \delta \pi^b(y)} j^b(y) .
\end{align}
Note that the first term of Eq.~\eqref{eq:gf2} is the effective action
that is proportional to $N_c$ as shown in Eq.~\eqref{eq:effac}. The
third term is known to be the one meson-loop contribution, which is
proportional to $1/N_c$. In the large $N_c$ limit, this is
suppressed. As shown in Eq.~\eqref{eq:gf2}, the inverse of the meson
propagator is expressed by 
\begin{align}
(K_{\pi}^{ab})^{-1} =\left. \frac{\delta^2 S_{\mathrm{eff}}}{\delta
  \pi^a(x) \delta \pi^b(y)}\right|_{\pi_c} .
\end{align}
Expanding $S_{\mathrm{eff}}$ with respect to the meson fields, we
obtain the inverse of the pion and kaon propagators in momentum space 
as follows~\cite{Blotz:1992pw,Watabe:1995ts}:  
\begin{align}
\frac1{Z_\pi(p^{2})} \frac1{\delta^4(0)} \frac{\delta^2
  S_{\mathrm{eff}}}{\delta \pi^a (p) \pi^a(-p)}\bigg{|}_{p^{2}=-m^{2}_{\pi}}
 & = (p^2 + m_\pi^2)\bigg{|}_{p^{2}=-m^{2}_{\pi}},\;\;\;a=1,\,2,\,3,\cr 
\frac1{Z_K(p^{2})} \frac1{\delta^4(0)} \frac{\delta^2
  S_{\mathrm{eff}}}{\delta \pi^a (p) \pi^a(-p)}\bigg{|}_{p^{2}=-m^{2}_{K}}
 & = (p^2 + m_K^2)\bigg{|}_{p^{2}=-m^{2}_{K}},\;\;\;a=4,\cdots,\,7.
\end{align}
The poles of the meson propagators yield the masses of the pion and
kaon 
\begin{align}
m_\pi^2 &= \frac{m_0 I_1(M, m_0)}{Z_\pi(p^2=-m_\pi^2)}  M =
          (139\,\mathrm{MeV})^2,\cr 
m_K^2 &= \frac{m_0I_1(M,m_0) + m_{\mathrm{s}}
        I_1(M,m_{\mathrm{s}})}{2Z_K(p^2=-m_K^2)}M + (m_0-m_{\mathrm{s}})^2 =
        (496\,\mathrm{MeV})^2 ,
\label{eq:mass}
\end{align}
where 
\begin{align}
I_1(M,m_0) = 4N_c \int \frac{d^4 k}{(2\pi)^4}
  \frac{M^2}{k^2+(M+m_0)^2}.   
\end{align}
The $m_{0}$ denotes the average current quark mass i.e.,
$m_{0}=(m_{u}+m_{d})/2$. $Z_\pi$ and $Z_K$ stand for the
renormalization constants for the pion and kaon wavefunctions, given
as  
\begin{align}
Z_\pi(p^2) &= 2N_c \int  \frac{d^4 k}{(2\pi)^4}
\frac{M^2}{k^2+(M+m_0)^2}\frac{M^2}{(k+p)^2 + (M+m_0)^2},\cr
Z_K(p^2) &= 2N_c \int  \frac{d^4 k}{(2\pi)^4}
\frac{M^2}{k^2+(M+m_0)^2}\frac{M^2}{(k+p)^2 + (M+m_s)^2}.
\end{align}
Since $I_1$ and the renormalization constants are divergent, we use
the proper-time regularizations in this
work~\cite{Blotz:1992pw,Watabe:1995ts}. 

The meson decay constants $f_\pi$ and $f_K$ are defined as the
transition from the meson state to the vacuum through the axial-vector
current 
\begin{align}
\langle 0| A_\mu^a(x) | \phi^b(q)\rangle  = i q_\mu f_\phi e^{-iq\cdot
  x} \delta^{ab} ,
\end{align}
where $\phi^a$ denotes generically the pion or the kaon. The
axial-vector current is given by $A_\mu^a(x) =i\psi^\dagger
(x)\gamma_\mu\gamma_5\frac{\lambda^a}{2} \psi(x) $ in Euclidean space.
After some length calculations~\cite{Christov:1995vm,
  Blotz:1992pw,Watabe:1995ts}, we obtain the pion and kaon decay
constants   
\begin{align}
\label{eq:decay}
f_\phi (p) = \frac{\sqrt{2Z_\phi(p)}}{M} . 
\end{align}
Using Eqs.~\eqref{eq:mass} and \eqref{eq:decay}, we obtain the
Gell-Mann-Oakes-Renner relation
\begin{align}
f_\pi^2 m_\pi^2 = -m_0 i\langle \psi^\dagger (0) \psi(0)\rangle_0, 
\end{align}
where the quark condensate $i\langle \psi^\dagger (0)
\psi(0)\rangle_0$ is given by
\begin{align}
\langle i\psi^\dagger (0) \psi(0)\rangle_0 = -2 \frac{I_1(M,m_0)}{M}.  
\end{align}
The cut-off parameter $\Lambda$ and the average current quark mass
$m_0$ are fixed by the pion decay constant and the pion mass,
respectively. The dynamical quark mass $M$ is considered to be a free
parameter. We take its value to be $M=420$ MeV, with which baryonic
properties have been well reproduced~\cite{Christov:1995vm}. The
strange current quark mass $m_{\mathrm{s}}=150$ MeV yields the kaon
mass $m_K=496$ MeV. In the present work, we do not consider the
effects of explicit flavor SU(3) symmetry breaking, since their
effects are marginal.   

\subsection{Kaon cloud}
Once the parameters of the model were fixed in the mesonic sector, we
can proceed to the baryonic sector. The effective chiral action
$S_{\mathrm{eff}}$ in Eq.~\eqref{eq:effac} can be expressed in terms
of the one-body Dirac Hamiltonian 
\begin{align}
H = \gamma_4\gamma_k \partial_k + M\gamma_4 U^{\gamma_5} +
  m_0\gamma_4, 
\label{eq:ham}
\end{align}
of which the eigenenergies and eigenstates of a quark are given by
$H|n\rangle = E_n|n\rangle$. Having computed the nucleon correlation
function~(see Refs.~\cite{Diakonov:1997sj, Christov:1995vm} for
details), we obtain the classical mass as 
\begin{align}
M_{\mathrm{cl}} = N_c E_{\mathrm{val}}[\pi^a] +E_{\mathrm{sea}}[\pi^a],  
\end{align}
where $E_{\mathrm{val}}$ denotes the energy given by the $N_c$ valence
quarks filled in the lowest upper Dirac level, which yields the state
with baryon number one. $E_{\mathrm{sea}}$ stands for the energy that
is required for the pion mean field to be created. This is just the
sum of the energies of the quarks filled in the lower Dirac continuum.   
Note that both the energies $E_{\mathrm{val}}$ and $E_{\mathrm{sea}}$
are given as the functionals of the pion field.  
Using the classical equation of motion for the pion mean field, we can
minimize the energy of the classical nucleon or the chiral soliton.  
The final solution of the pion mean field is expressed as
\begin{align}
\pi_c^a(\bm{x}) = n^a P(r),   
\end{align}
where $P(r)$ designates the profile function of the chiral soliton and
$n^a$ is a normal vector in isospin space, defined as
$n^a=x^a/\|\bm{x}\|$. In SU(3), we embed this stationary solution into
SU(3) as follows:
\begin{align}
U(\bm{x}) := \begin{pmatrix}
\exp[i\bm{n}\cdot \bm{\tau} P(r)] & 0 \\
0 & 1
\end{pmatrix} .
\label{eq:su3}
\end{align}
$m_0$ in Eq.~\eqref{eq:ham} plays an essential role in producing the
correct proper Yukawa-type asymptotic behavior of the profile
function
\begin{align}
P(r) = \alpha \exp(-\overline{m}_{\pi} r) \frac{1+\overline{m}_\pi
  r}{r^2},  
\label{eq:Prof_taol}
\end{align}
where $\alpha$ denotes a constant that governs the strength of the
profile function and $\overline{m}_\pi$ represents the generic meson
mass that produces a required meson tail. For instance, the
$\overline{m}_\pi = m_\pi=139$ MeV corresponds to the pion cloud
whereas $\overline{m}_\pi = m_K=496$ MeV produces the proper kaon
cloud. As mentioned in the Introduction, however, the
$\Omega$ is not coupled to the pion but to the kaon. Unfortunately,
any chiral solitonic approach cannot take the pion and kaon mean
fields into account separately. Since we quantize the SU(3) soliton in 
Eq.~\eqref{eq:su3} by rotating it slowly in the coordinate and flavor 
spaces, both the pion and kaon fields have the same pion tail. To
investigate properties of the $\Omega^-$ baryon, it is natural to
consider only the kaon mean field. Of course, this is a rather
phenomenological approach but is a necessary one to describe
$\Omega^-$ more physically. 
Thus, we increase the value of $m_0$ such
that the proper kaon cloud is produced, i.e.,
$\overline{m}_{\pi}=m_K=496$ MeV in
Eq.~\eqref{eq:Prof_taol}.  Once we obtain the profile function with
the kaon cloud, we can compute various properties of $\Omega^-$. As
will be shown soon, the chiral soliton with the kaon cloud describes
the EM and axial-vector form factors far better than that with the
pion cloud, when the results are compared with those from lattice QCD.  
Since we are mainly interested in the effects of the kaon cloud, we
will compute all the observables, imposing flavor SU(3) symmetry
without its explicit breaking considered. 

One may argue that the effects of flavor SU(3) symmetry breaking are
too large to be ignored as shown in the Skyrme model with the
bound-state approach~\cite{Callan:1987xt} and Yabu-Ando’s
approach~\cite{Yabu:1987hm}. In fact, the Yabu-Ando’s approach has
been examined within the present framework~\cite{Blotz:1992pw}, where
they have computed the mass spectrum of the hyperons. The results
showed clearly that the higher-order corrections
($\mathcal{O}(m_s^3)$) are negligible, since they are rather similar
to those obtained by the perturbative approach. This indicates that
the effects of flavor SU(3) symmetry breaking are rather small. In
fact, we found that these effects turned out to be at most 10~\% for
almost all baryonic observables~\cite{Christov:1995vm}. So, we will
only consider the flavor SU(3) symmetric case.

\section{Electromagnetic and axial-vector form factors
  of the  $\Omega^-$  baryon} 
We first compute the EM form factors of the $\Omega^-$ baryon. Since
the general formalism was presented in Ref.~\cite{Kim:2019gka} in
detail, we briefly recapitulate it. 
The matrix element of the EM current in Euclidean space is defined as 
\begin{align}
  \label{eq:LHcurrent}
J^\mu (x) := i \psi^\dagger (x) \gamma^\mu \hat{\mathcal{Q}} \psi(x),  
\end{align}
where $\psi(x)$ denotes the quark field. The charge operator of the
quarks $\hat{\mathcal{Q}}$ is written in terms of the flavor SU(3)
Gell-Mann matrices $\lambda_3$ and $\lambda_8$ 
\begin{align}
 \label{eq:chargeOp}
\hat{\mathcal{Q}} =
  \begin{pmatrix}
   \frac23 & 0 & 0 \\ 0 & -\frac13 & 0 \\ 0 & 0 & -\frac13
  \end{pmatrix} = \frac12\left(\lambda_3 + \frac1{\sqrt{3}}
                                                  \lambda_8\right).
\end{align}
The matrix elements of the EM current between the $\Omega^-$ baryons
can be parametrized by four form factors $F_i$ ($i=1,\cdots,4$) as
follows:  
\begin{align}
\langle \Omega^-(p',s) |  J^{\mu}(0) | \Omega^-(p,s) \rangle 
&= - \overline{u}^{\alpha}(p',s) \left[ \gamma^{\mu} \left \{
  F_{1}(q^2) \eta_{\alpha \beta} + F_{3}(q^2) \frac{ q_{\alpha} q_{\beta}
  }{4M_{\Omega}^{2}}  \right \}\right. \cr
&\hspace{2cm} \left.  + \,i\frac{\sigma^{\mu \nu} q_{\nu}}{2M_{\Omega}}
  \left \{ F_{2}(q^2) \eta_{\alpha \beta} + F_{4} (q^2)\frac{q_{\alpha}
  q_{\beta}}{4 M_{\Omega}^2}  \right \}  \right ]{u}^{\beta}(p,s), 
\label{eq:MatrixEl1}
\end{align}
where $M_{\Omega}$ stands for the mass of the $\Omega^-$ baryon in the 
decuplet. $q_{\alpha}$ represents the momentum transfer 
$q_{\alpha}=p_{\alpha}'-p_{\alpha}$ and its square
is given by $q^2=-Q^2$ with $Q^2 >0$. $u^\alpha (p,\,s)$ designates
the Rarita-Schwinger spinor, carrying the momentum $p$ and the spin
component $s$ projected along the direction of the momentum.  

The multipole EM form factors can be expressed in terms of $F_i$
in Eq.~\eqref{eq:MatrixEl1}
\begin{align}
&G_{E0}(Q^{2}) = \left(1 + \frac{2}{3} \tau\right)[F_{1}-\tau F_{2}] -
  \frac{1}{3} \tau (1+ \tau) [F_{3} - \tau F_{4}], \cr 
&G_{E2}(Q^{2}) = [F_{1}-\tau F_{2}]- \frac{1}{2} (1+ \tau)
  [F_{3} - \tau F_{4}] , \cr 
&G_{M1} (Q^{2}) = \left(1+\frac{4}{5} \tau\right)[F_{1}+F_{2}]  -
  \frac{2}{5}  \tau (1+ \tau)[F_{3}+F_{4}], \cr 
&G_{M3}(Q^{2}) =  [F_{1}+F_{2}] -\frac{1}{2}(1+\tau)
  [F_{3}+F_{4}], 
\end{align}
where $\tau=Q^2/4M_{\Omega}^2$. These four form factors are called, 
respectively, the electric or Coulomb monopole (E0), magnetic dipole
(M1), electric quadrupole (E2), and magnetic octupole (M3) form
factors. At $Q^2=0$, the values of these form factors are reduced to
the charge, the magnetic dipole moment, the electric quadrupole
moment, and the magnetic octupole moment, respectively
\begin{align}
e_{\Omega} &= e G_{E0} (0) ,\cr
\mu_{\Omega} &= G_{M1}(0) \left(\frac{M_N}{M_\Omega}\right)
                 \mu_N,\cr
Q_{\Omega} &= \frac{e}{M_{\Omega}^2} G_{E2}(0),\cr
O_{\Omega} &= \frac{e}{M_{\Omega}^3} G_{M3}(0).
\label{eq:atq0}
\end{align}
Note that the electric quadrupole moment is proportional to $1/N_c$
and the magnetic octupole moment is of order $1/N_c^2$. Thus, M3  
vanishes when we consider the corrections to order $1/N_c$. 

In the Breit frame, i.e., $\bm{p'}=-\bm{p}=\bm{q}/2$, the electric and
magnetic parts of the multipole form factors are related to the
temporal and spatial components of the EM current, respectively 
\begin{align}
G_{E0}(Q^{2}) &= \int \frac{d \Omega_{q}}{4 \pi}\langle
                \Omega^-(p',3/2) | 
  J^{0}(0) | \Omega^- (p,3/2) \rangle,  \cr 
G_{E2}(Q^{2}) &= -\int d \Omega_{q} \sqrt{\frac{5}{4\pi}} \frac{3}{2}
                \frac1{\tau} \langle \Omega^-(p',3/2) | Y^{*}_{20}  
  (\Omega_{q}) J^{0}(0) | \Omega^- (p,3/2) \rangle, \cr 
G_{M1}(Q^{2})&= \frac{3 M_\Omega}{4\pi}\int 
\frac{d\Omega_{q}}{ iQ^{2}} q^{i} \epsilon^{ik3} 
\langle \Omega^- (p',3/2) | J^{k}(0) | \Omega^-   (p,3/2) \rangle,  \cr
G_{M3}(Q^{2}) &=-\frac{35M_{\Omega}}{8}\sqrt{\frac{5}{\pi}} \int 
  \frac{d\Omega_{q}}{ i  Q^{2} \tau}q^{i}\epsilon^{ik3}\langle
  \Omega^- (p',3/2) |  \left( 
Y^{*}_{20}(\Omega_{q}) + \sqrt{\frac{1}{5}}Y^{*}_{00}(\Omega_{q}) 
                  \right) 
J^{k}(0) | \Omega^- (p,3/2) \rangle.
\label{eq:matel6}  
\end{align}
The matrix elements of the EM current can be computed within the
framework of the $\chi$QSM, which was already done in
Ref.~\cite{Kim:2019gka}, with the rotational $1/N_c$ and linear
$m_{\mathrm{s}}$ corrections considered. For detailed
calculations, we refer to Ref.~\cite{Kim:2019gka}. For convenience, we
compile the expressions for the EM form factors of the $\Omega$ baryon
in Appendix~\ref{app:a}.  

We can derive the axial-vector form factors of the $\Omega^-$ baryon
in a similar manner. Since $\Omega^-$ is an isosinglet baryon, the
triplet components of the $\Omega^-$ axial-vector form factors
vanish. The axial-vector current of the quark field is defined as 
\begin{align}
  \label{eq:Axcurrent}
A_\mu^0:= i\psi^\dagger (x) \gamma_\mu
  \gamma^{5} \psi(x), \;\;\;
 A_\mu^8 (x) := i\psi^\dagger (x) \gamma_\mu
  \gamma^{5}\frac{\lambda^{8}}{2} \psi(x).    
\end{align}
The matrix elements of the axial-vector currents for the
$\Omega^-$ baryon can again be parametrized by four different real
form factors as follows:  
\begin{align}
\langle \Omega^-(p',s) | A_{\mu}^0 (0) | \Omega^- (p,s) \rangle 
&= - \overline{u}^{\alpha}(p',s) \left[ \gamma_{\mu} \left \{
  g_{1}^{(0)}(q^2) \eta_{\alpha \beta} + h_{1}^{(0)}(q^2) \frac{
  q_{\alpha} q_{\beta} }{4M_{\Omega}^{2}}  \right \}\right. \cr
&\hspace{2cm} \left.  + \,\frac{q^{\mu}}{2M_{\Omega}}
  \left \{ g_{3}^{(0)}(q^2) \eta_{\alpha \beta} +
  h_{3}^{(0)}(q^2)\frac{q_{\alpha} 
  q_{\beta}}{4 M_{\Omega}^2}  \right \}  \right
  ]\gamma^{5} u^{\beta}(p,s), \cr
\langle \Omega^-(p',s) | A_{\mu}^8 (0) | \Omega^-(p,s) \rangle 
&= - \overline{u}^{\alpha}(p',s) \left[ \gamma_{\mu} \left \{
  g_{1}^{(8)}(q^2) \eta_{\alpha \beta} + h_{1}^{(8)}(q^2) \frac{
  q_{\alpha} q_{\beta}   }{4M_{\Omega}^{2}}  \right \}\right. \cr
&\hspace{2cm} \left.  + \,\frac{q^{\mu}}{2M_{\Omega}}
  \left \{ g_{3}^{(8)}(q^2) \eta_{\alpha \beta} +
  h_{3}^{(8)}(q^2)\frac{q_{\alpha} 
  q_{\beta}}{4 M_{\Omega}^2}  \right \}  \right
  ]\frac{\gamma^{5}}{2}{u}^{\beta}(p,s).
\label{eq:MatrixEl2}
\end{align}

Since the axial-vector form factors $g_1^{(0,\,8)}$ and $g_3^{(0,\,8)}$ are
the most important ones among the axial-vector form factors, we will
concentrate on them in the present work. Moreover, they are directly 
related to the strong coupling constants for the $\eta \Omega \Omega$
and $\eta' \Omega\Omega$ vertices~\cite{Yang:2018idi}.
In the Breit frame, the form factors defined in Eq
~\eqref{eq:MatrixEl2} are expressed in terms of the spatial parts of
the axial-vector current projected by the spherical basis vectors
$\bm{e}_{n}$ 
\begin{align}
g^{(0,\,8)}_{1}(Q^2) &= -\sqrt{\frac32}\frac{ M_{\Omega}}{E_{\Omega}}
\langle \Omega^- (p',3/2) | \bm{e}_{1} \cdot
  \left(\bm{A}^{0} (0), 2\bm{A}^{8} (0) \right)
    | \Omega^- (p, 1/2)\rangle , \cr  
g^{(0,\,8)}_{3}(Q^2)&= - \frac{4M^{2}_{\Omega}}{Q^{2}} 
  \left[ \langle \Omega^- (p',3/2) | \bm{e}_{0} \cdot 
  \left(\bm{A}^{0} (0), 2\bm{A}^{8} (0) \right) 
  |\Omega^- (p,3/2) \rangle \right. \cr
&\left. \hspace{2cm} - \,\sqrt{\frac{3}{2}}
                      \frac{M_{\Omega}}{E_{\Omega}}   
  \langle \Omega^- (p',3/2) | \bm{e}_{1} \cdot 
\left(\bm{A}^{0} (0), 2\bm{A}^{8} (0) \right)
   |\Omega^- (p,1/2) \rangle \right] ,
\label{eq:MatrixEl3}
\end{align}
where $E_{\Omega}$ corresponds to the energy of the corresponding
baryon, i.e. $E_{\Omega}=\sqrt{M^{2}_{\Omega} +Q^{2}/4}$, and the
spherical unit vectors $\bm{e}_{n}$ are
expressed in terms of the Cartesian basis vectors
$\bm{e}_{0}=\hat{\bm{z}}$, $\bm{e}_{1}=
-(\hat{\bm{x}}+i\hat{\bm{y}})/ \sqrt{2}$, $\bm{e}_{-1}=
(\hat{\bm{x}}-i\hat{\bm{y}})/ \sqrt{2}$. 
The matrix elements of the projected currents $\bm{e}_{i} \cdot
 \left(\bm{A}^{0} (0), 2\bm{A}^{8} (0) \right)$ 
can be computed within the framework of the
$\chi$QSM. The detail formalism can be found in Ref.~\cite{Jun:2020lfx}. 
We list the expressions for the EM form factors of the $\Omega$ baryon 
in Appendix~\ref{app:b}.  
\section{Results and discussion}
\begin{figure}[htp]
\includegraphics[scale=0.27]{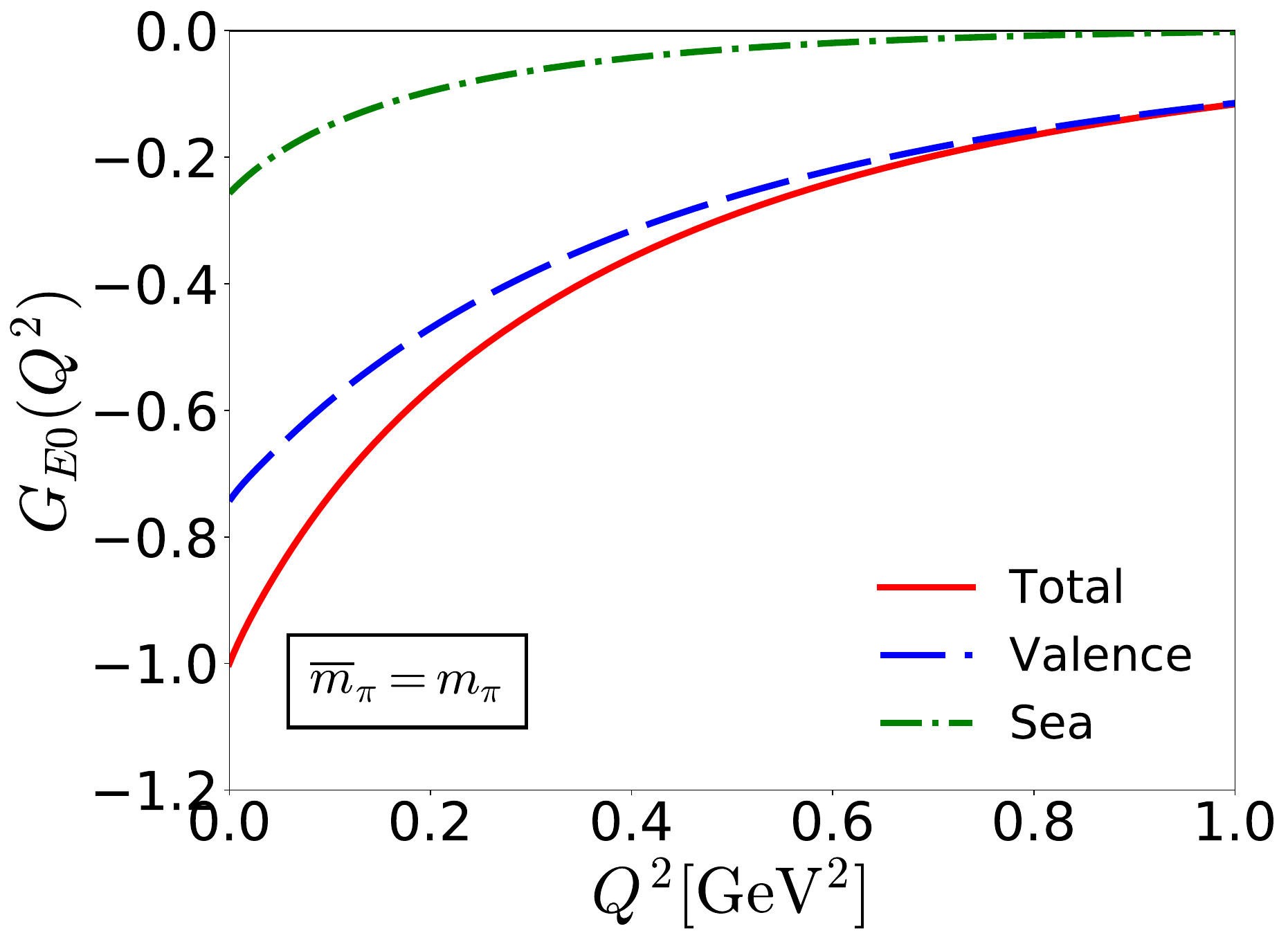}
\includegraphics[scale=0.27]{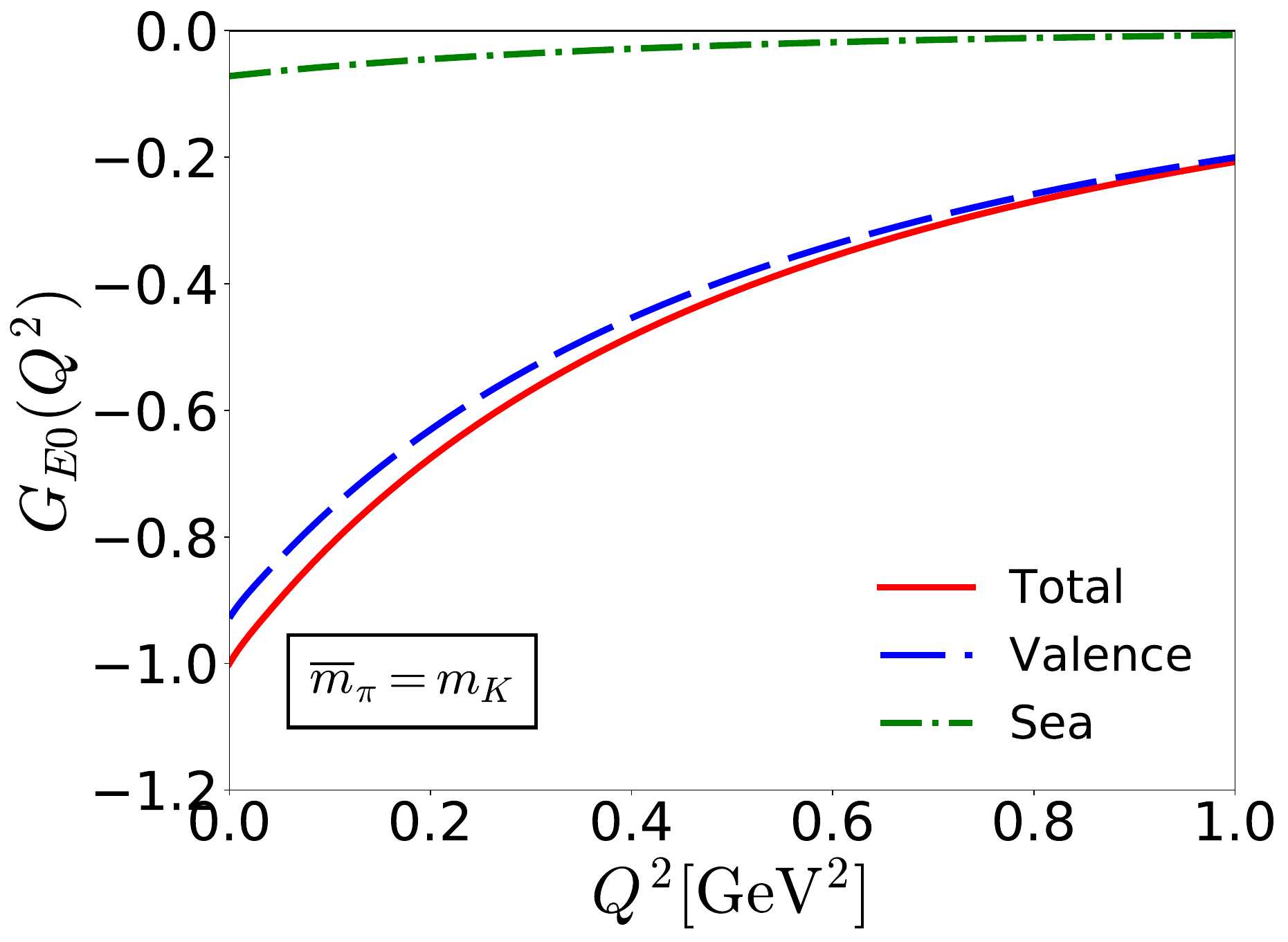}
\includegraphics[scale=0.27]{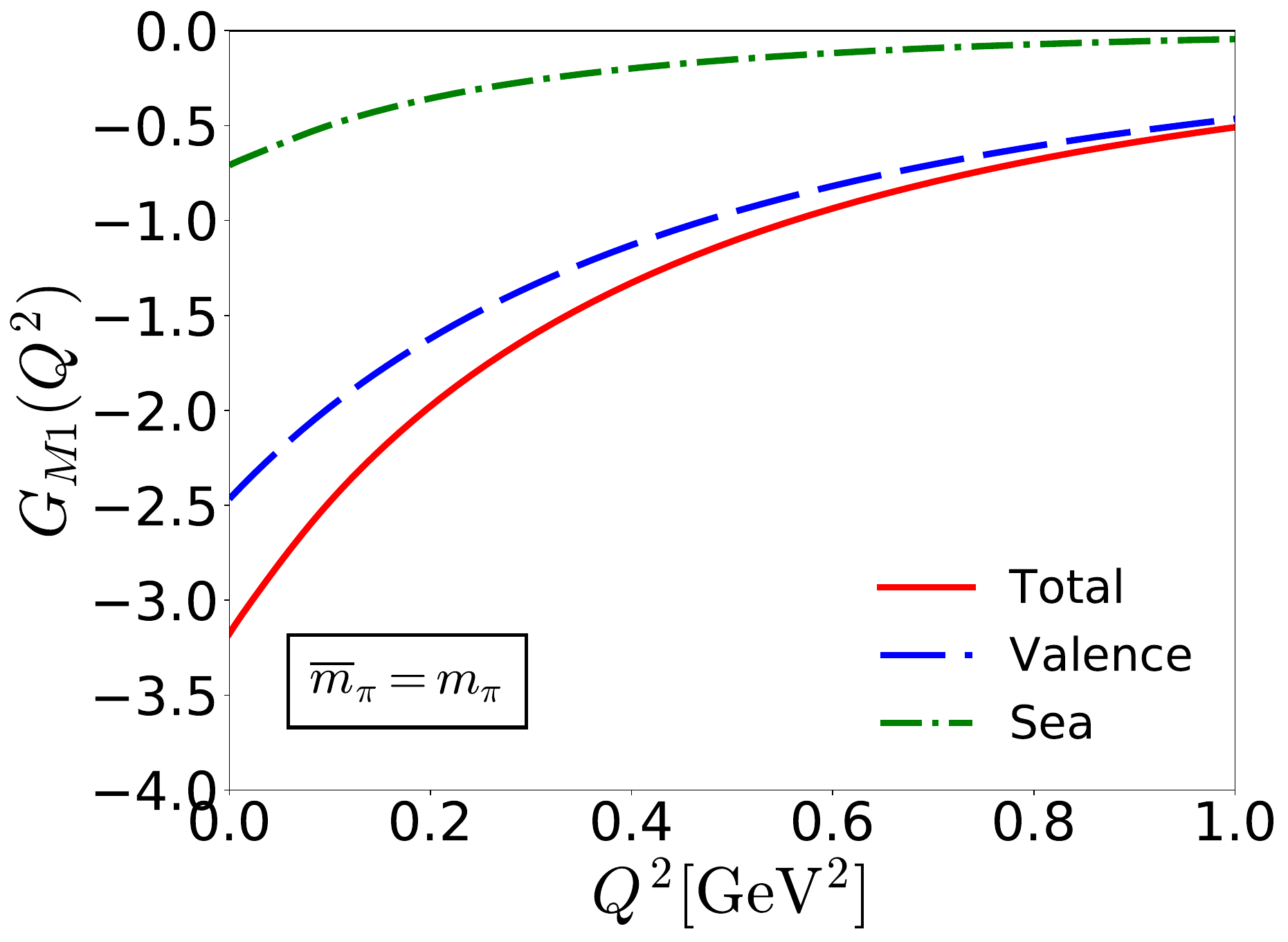}
\includegraphics[scale=0.27]{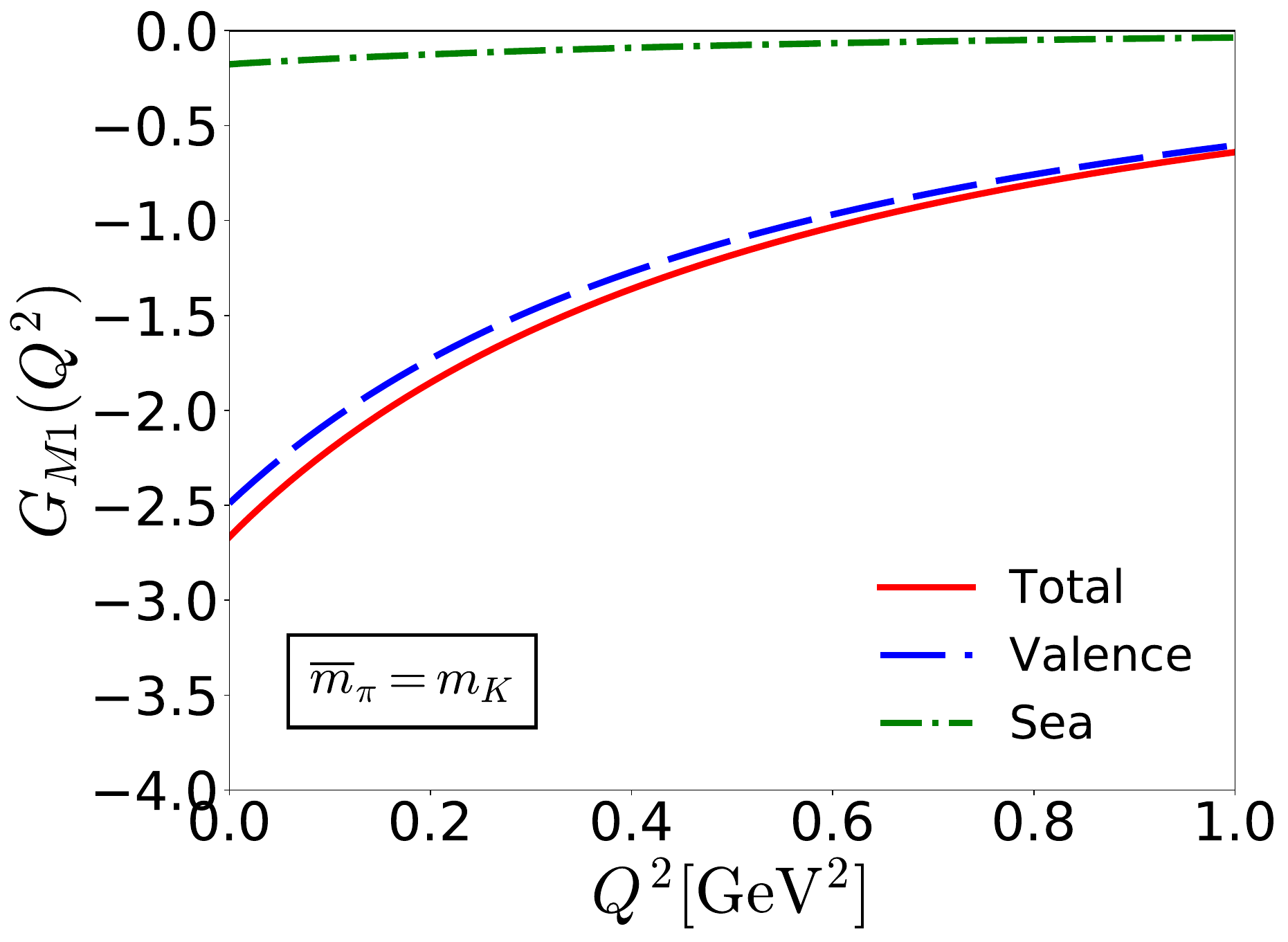}
\includegraphics[scale=0.27]{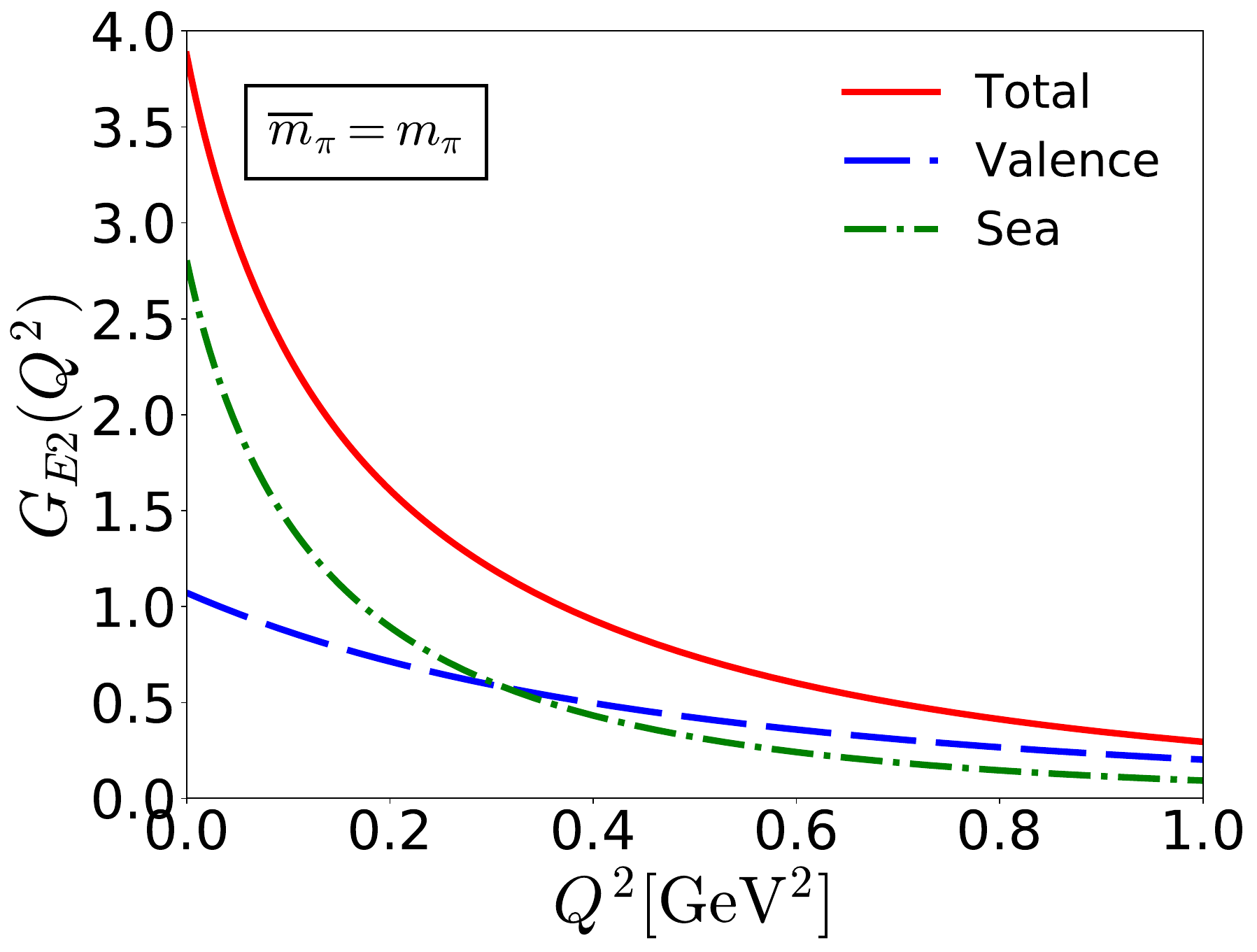}
\includegraphics[scale=0.27]{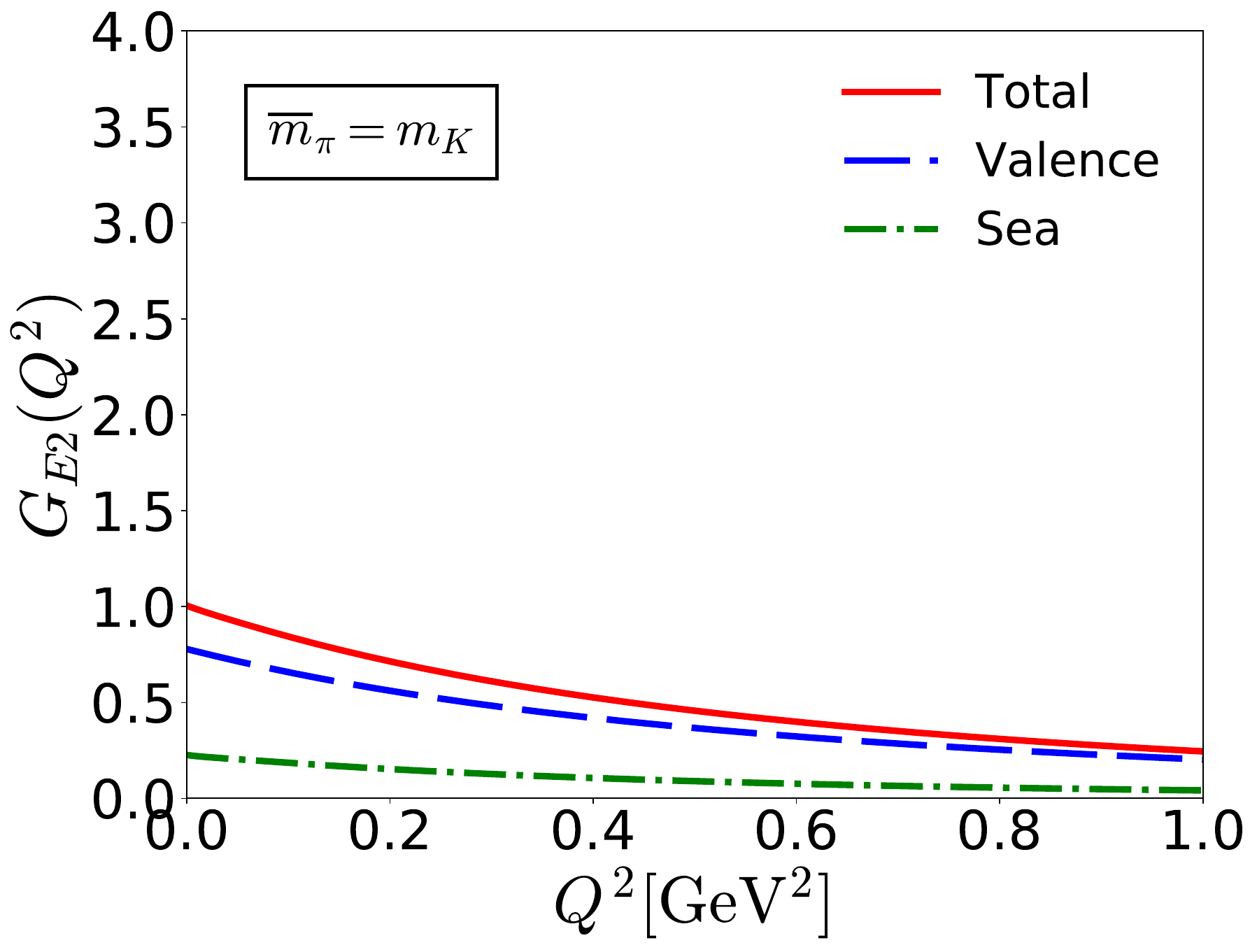}
\caption{Numerical results for the E0, M1, E2 form factors of the
  $\Omega^-$ baryon with both the pion and kaon clouds considered. In
  the left panels those with the pion cloud are drawn whereas in the
  right panels those with the kaon cloud are depicted. The dashed and
  dot-dashed curves exhibit the valence-quark
  (level-quark) and sea-quark (Dirac continuum) contributions,
  respectively. The solid curves show the total contributions. }
\label{fig:1}
\end{figure}
We first present the results for the E0, M1, and E2 form factors of
the $\Omega^-$ baryon, drawing separately the valence- and sea-quark 
contributions. We already expect that the sea-quark contributions will
be greatly changed by replacing the pion cloud with the kaon one. 
In the upper-left (right) panel of Fig.~\ref{fig:1}, we depict the
numerical results for the E0 form factors with the pion (kaon) cloud
employed. Since the E0 form factor of $\Omega^-$ at $Q^2=0$ should be
the same as its charge as shown in Eq.~\eqref{eq:atq0} because of the
U(1) gauge symmetry, one can see that replacing the pion cloud with the
kaon one enhances the valence-quark contribution but suppresses that
of the sea-quark. We find similar features for the M1 and
E2 form factors by observing the results presented respectively in the
middle and lower panels of Fig.~\ref{fig:1}. In particular, the
sea-quark contribution to the E2 form factor is drastically reduced
when the pion cloud is replaced with the kaon cloud. As already
discussed in Ref.~\cite{Kim:2019gka}, the E2 form factors of the
baryon decuplet are in general very sensitive to the tail. This
indicates that the deformation of a baryon with spin 3/2 is governed
by the sea-quark contribution or the meson clouds. On the other hand,
the valence-quark contributions are not much influenced by changing
the meson clouds.

\begin{figure}[htp]
\includegraphics[scale=0.27]{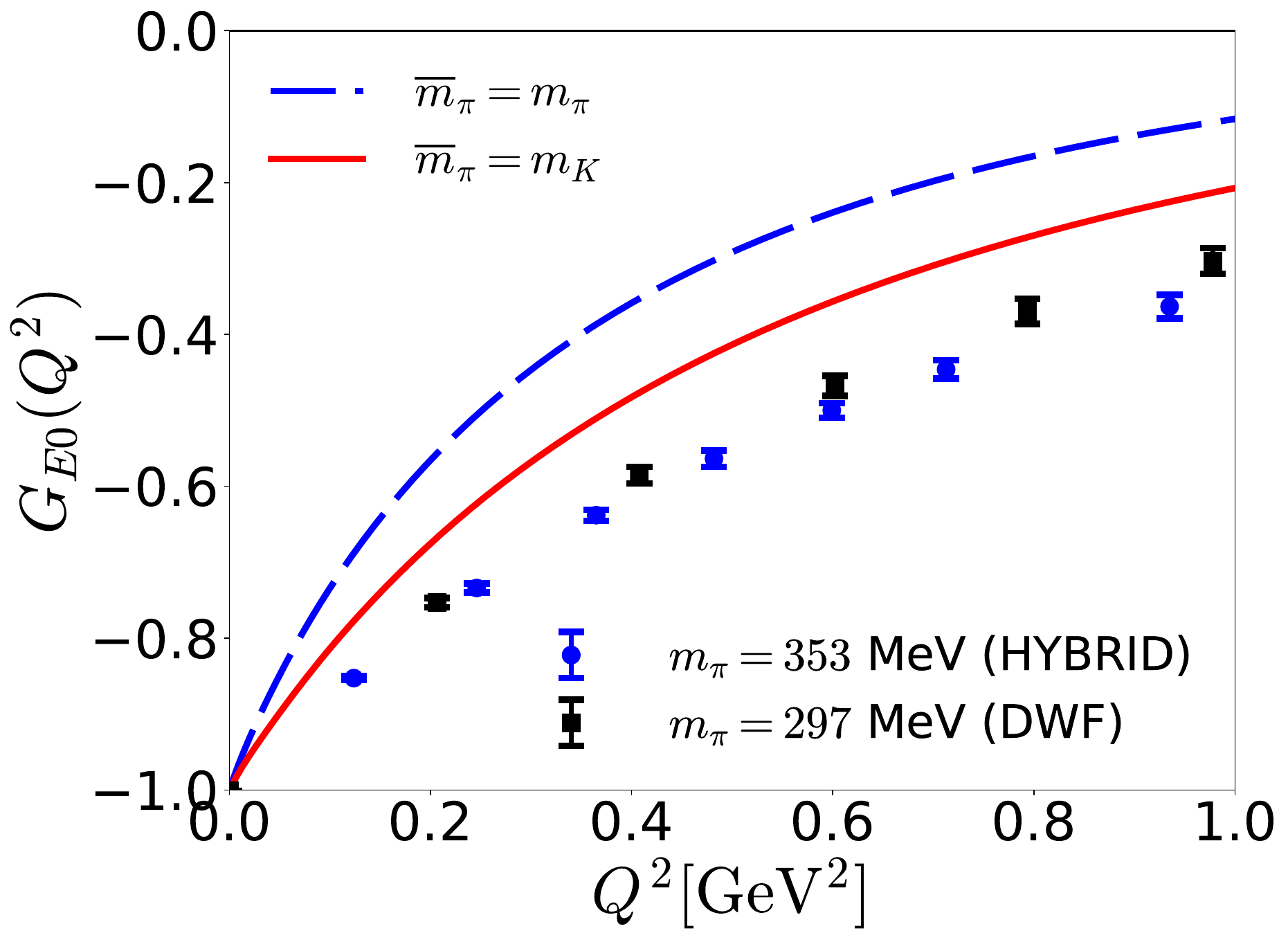}
\includegraphics[scale=0.27]{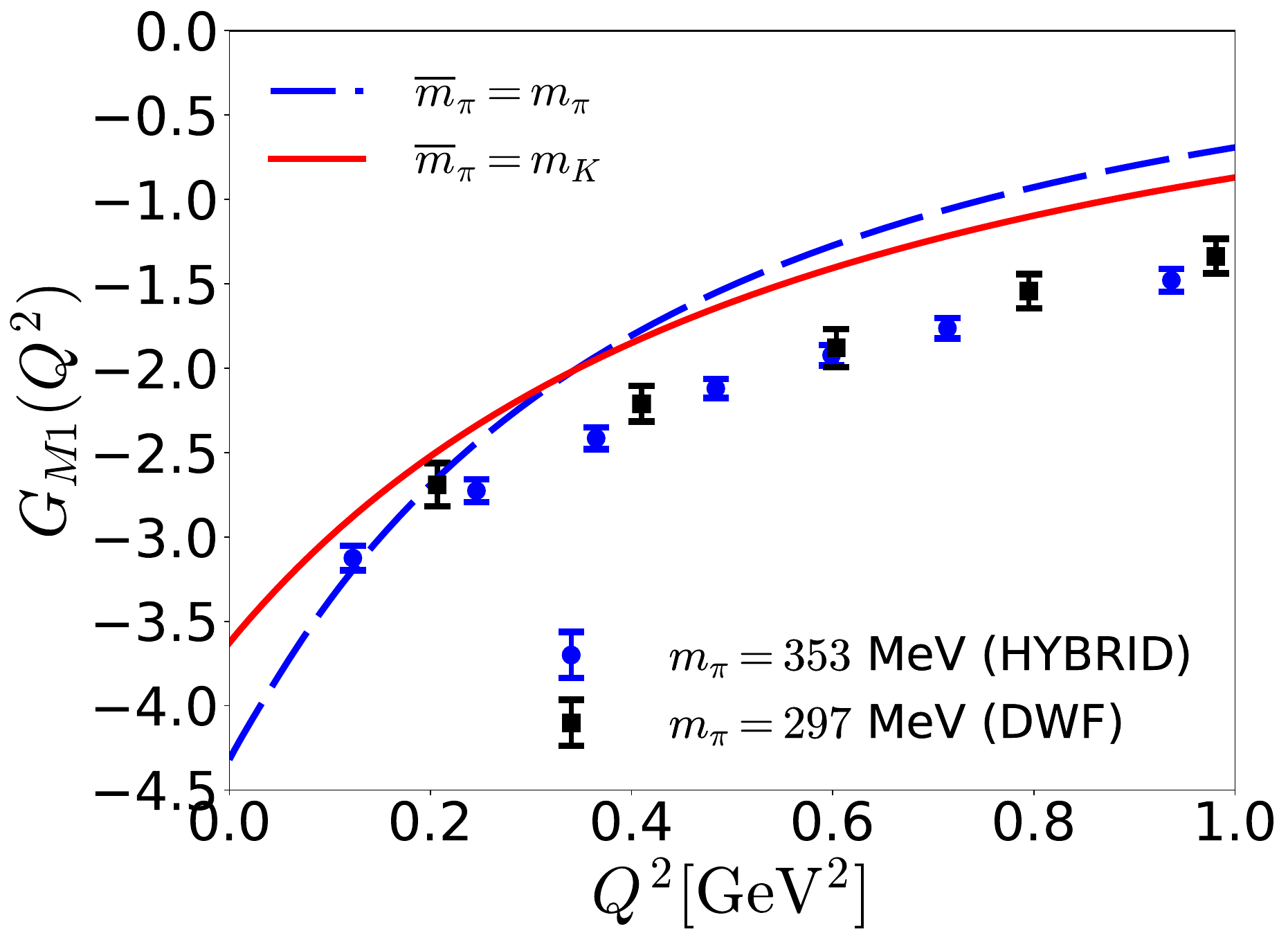} 
\includegraphics[scale=0.27]{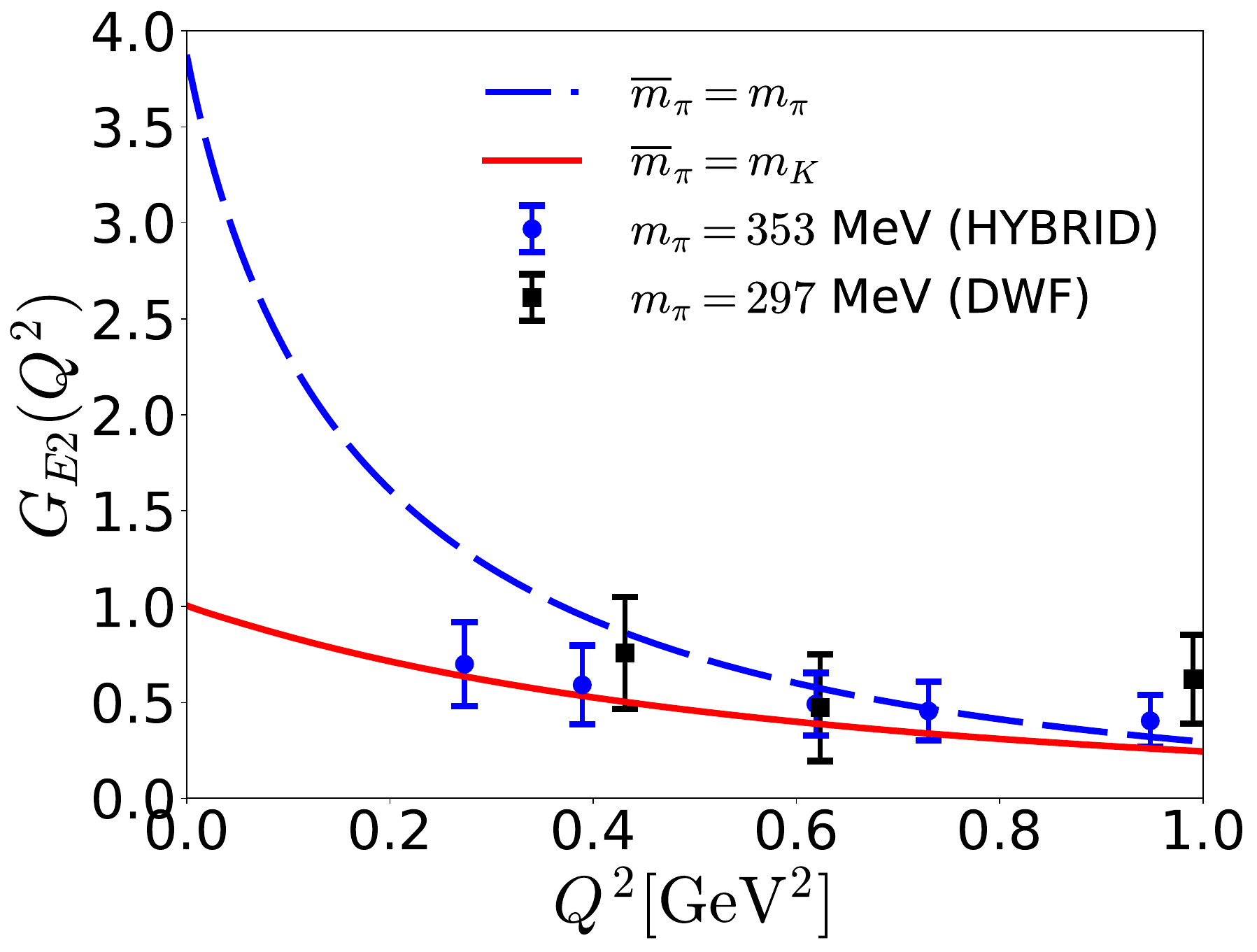}
\caption{Comparison of the present results for the EM form factors of
  the $\Omega^-$ baryon with those from lattice data. The solid curves
draw the results with the kaon cloud whereas the dashed ones depict
those with the pion cloud. The lattice data are taken from
Ref.~\cite{Alexandrou:2010jv}. The magnetic form factor is given in
units of the soliton magneton, of which the reason is explained in the
text.}    
\label{fig:2}
\end{figure}
In Fig.~\ref{fig:2}, we compare the present results for the EM form
factors of $\Omega^-$ with those from lattice
QCD~\cite{Alexandrou:2010jv}. In the upper left panel of
Fig.~\ref{fig:2}, we draw the numerical results for the E0 form factor
of the $\Omega^-$. The result with the kaon cloud falls off more
slowly than that with the pion cloud and is in better agreement with
the lattice data, compared with that with the pion cloud. 
It is well known that the lattice data on the EM form factors of the
nucleon~\cite{Capitani:2015sba, Hansen:2016fzj} with the unphysical
pion mass fall off more slowly than the experimental data as $Q^2$
increases. Even the physical pion mass is used, the lattice results
for the EM form factors of the nucleon still overestimate the
experimental data~\cite{Alexandrou:2017ypw}. Thus, it is natural to
observe that the present results for the E0 form factor of the
$\Omega^-$ baryon fall off faster than the lattice data, as shown in
the upper left panel of Fig.~\ref{fig:2}. However, the result with the
kaon cloud is markedly closer to the lattice data than those with the
pion one.   

In the upper right panel of Fig.~\ref{fig:2}, we depict the results
for the M1 form factor of $\Omega^-$. Before we discuss them, we want
to mention that the magnetic dipole moments of the SU(3) baryons are
given in terms of the soliton magneton instead of the nuclear
magneton. There are two reasons for doing this. Firstly, while the
mass differences of the baryon octet and decuplet and even singly
heavy baryons are well reproduced within the
$\chi$QSM~\cite{Blotz:1992pw, Christov:1995vm, Kim:2018xlc,
  Kim:2019rcx}, the absolute values of these masses are always
overestimated. This is a well-known problem in any chiral solitonic
approaches. Secondly, the magnetic dipole moments of the baryon octet
and decuplet are always underestimated by about 30~\%. Thus, it is
theoretically consistent and empirically plausible to express the
values of the magnetic dipole moments in units of the soliton
magneton, which improves theoretical results for the magnetic dipole
moments in comparison with the experimental data. 
As shown in the upper right panel of Fig.~\ref{fig:2}, the result for
the M1 form factors of $\Omega^-$ with the kaon cloud again is in
better agreement with the lattice data, compared with that with the
pion one. 

The lower panel of Fig.~\ref{fig:2} illustrates how the kaon
cloud suppresses the E2 form factor of $\Omega^-$. It is remarkable
that the kaon cloud reduces the magnitude of the E2 form factor almost
by factor 4 and the numerical result with the kaon cloud is in far
better agreement with the lattice data, compared to that with the pion
cloud. As already seen in Ref.~\cite{Kim:2019gka}, the sea-quark
contribution or the meson-cloud effect is dominant over that of the 
valence quarks. This is natural, since the E2 form factor measures how
much the baryon with spin 3/2 is deformed from the spherical
shape. This implies that the $\Omega^-$ baryon is less deformed than
the $\Delta$ isobar, since it is energetically easier to create the
pion cloud than the kaon one. Comparing the results with the lattice
data, we observe that the numerical result with the kaon cloud is
indeed in good agreement with them. On the other hand,
that with the pion cloud  deviates from the lattice data in smaller
$Q^2$ regions.  
 
\begin{table}[htp]
\renewcommand{\arraystretch}{1.7}
\caption{Numerical results for the electromagnetic observables 
  in comparison with the lattice data~\cite{Alexandrou:2016xok}.
Those for the magnetic dipole moment of the $\Omega^-$ baryon are
compared with the experimental data~\cite{Zyla:2020zbs}.} 
\label{tab:1}
{\setlength{\tabcolsep}{6pt}
 \begin{tabular}{ c | c c c }
  \hline 
  \hline 
 & $\langle r^{2}
   \rangle_{E}$\,(fm$^{2}$) & $
 \mu$\,($\mu_{N}$)   & $G_{E2}(0)$ \\ 
  \hline 
$\bar{m}_{\pi}=m_{\pi}$ & $0.83$ & $-2.48$ & $3.88$ \\  
$\bar{m}_{\pi}=m_{K}$ & $0.51$ & $-2.04$ & $1.00$ \\  
Exp~\cite{Zyla:2020zbs}
 & $-$ & $-2.02(5)$ & $-$ \\
LQCD~\cite{Alexandrou:2010jv}
 & $0.348(52)$ & $-1.875(399)$ & $ 0.898(60)$ \\
 \hline
 \hline
\end{tabular}}
\end{table}
\setlength{\tabcolsep}{2pt}
\renewcommand{\arraystretch}{1.5}
\begin{table}[htp]
\caption{Magnetic dipole moment of the $\Omega^{-}$ baryon in
  comparison with the results from lattice
QCD~\cite{Leinweber:1992hy,Aubin:2008qp,Boinepalli:2009sq},  
the relativistic quark model~\cite{Schlumpf:1993rm},
 next-to-leading-order HB$\chi$PT~\cite{Butler:1993ej}, 
large $N_{c}$~\cite{Luty:1994ub}, QCD sum rules~\cite{Lee:1997jk},
the chiral quark model~\cite{Wagner:2000ii},
covariant $\chi$PT~\cite{Geng:2009ys}, $\chi$PT~\cite{Li:2016ezv} 
and the experimental
data~\cite{Zyla:2020zbs}.} 
\label{tab:2}
 \begin{tabular}{c | c c c c c c c c c c c c c c c  } 
  \hline 
    \hline 
  & $\overline{m}_{\pi}=m_{\pi}$ & $\overline{m}_{\pi}=m_{K}$ 
& \cite{Zyla:2020zbs} & \cite{Leinweber:1992hy} 
& \cite{Boinepalli:2009sq} & \cite{Aubin:2008qp} 
& \cite{Butler:1993ej} & \cite{Geng:2009ys} 
& \cite{Li:2016ezv} & \cite{Luty:1994ub} 
& \cite{Schlumpf:1993rm} & \cite{Wagner:2000ii}& \cite{Lee:1997jk}\\
  \hline 
$ \mu_{\Omega^{-}}(\mu_{N})  $ & $-2.48$ & $-2.04$ 
& $-2.02(5)$ & $-1.73(22) $ & $-1.70(7)$ & $-1.93(8)$ 
& $-1.94(22)$ &  $-2.02$ & $-2.02(5)$  & $-1.94$ 
& $-2.35$ & $-2.13$ &  $-1.49(45)$\\  
 \hline 
 \hline
\end{tabular}
\end{table}
\setlength{\tabcolsep}{5pt}
\renewcommand{\arraystretch}{1.5}
\begin{table}[htp]
\caption{Electric quadrupole moment of the $\Omega^{-}$ baryon in
  comparison with the quark model~\cite{Krivoruchenko:1991pm},
  HB$\chi$PT~\cite{Butler:1993ej}, the Skyrme model~\cite{Oh:1995hn},
  large $N_{c}$~\cite{Buchmann:2002et}, the chiral quark
  model~\cite{Wagner:2000ii}, the QCD sum 
  rules~\cite{Azizi:2008tx,Aliev:2009pd}.} 
\label{tab:3}
 \begin{tabular}{c |  c c c c c c c c c c  } 
  \hline 
    \hline 
  & $\overline{m}_{\pi}=m_{\pi}$ & $\overline{m}_{\pi}=m_{K}$ & 
\cite{Butler:1993ej} & \cite{Buchmann:2002et}  & 
\cite{Krivoruchenko:1991pm} & \cite{Oh:1995hn} & 
\cite{Wagner:2000ii} & \cite{Azizi:2008tx,Aliev:2009pd} \\
  \hline  
$ Q_{\Omega^{-}}(\mathrm{fm}^{2})  $ & $0.054$ & $0.014$ 
& $0.009(5)$ & $0.027$ & $0.028$ & $0$ & $0.026$ & $0.12(4)$\\  
 \hline 
 \hline
\end{tabular}
\end{table}
Table~\ref{tab:1} lists the numerical results for the charge radius,
magnetic dipole moment, and the value of the E2 form factor at $Q^2=0$
i.e., $G_{E2}(0)$ of the $\Omega^-$ baryon. Those with the kaon cloud
are prominently in better agreement with the lattice data than those
with the pion cloud. In the case of the magnetic dipole moment, we
also find that the result with the kaon cloud is in better agreement
with the result from lattice QCD. In Tables~\ref{tab:2}
and~\ref{tab:3}, we compare the present results respectively for the
magnetic dipole moment and the electric quadrupole moment of the
$\Omega^-$ baryon with those obtained from other approaches. As
already discussed in Ref.~\cite{Kim:2019gka}, the present numerical
result for the E2 moment turns out to be larger than
those from other works. We now can see that this discrepancy arises
from the fact that only when the kaon cloud is properly considered,
the E2 moment can be correctly reproduced. This emphasizes the
important role of the kaon cloud in describing the $\Omega^-$ baryon.   

\begin{figure}[htp]
\includegraphics[scale=0.27]{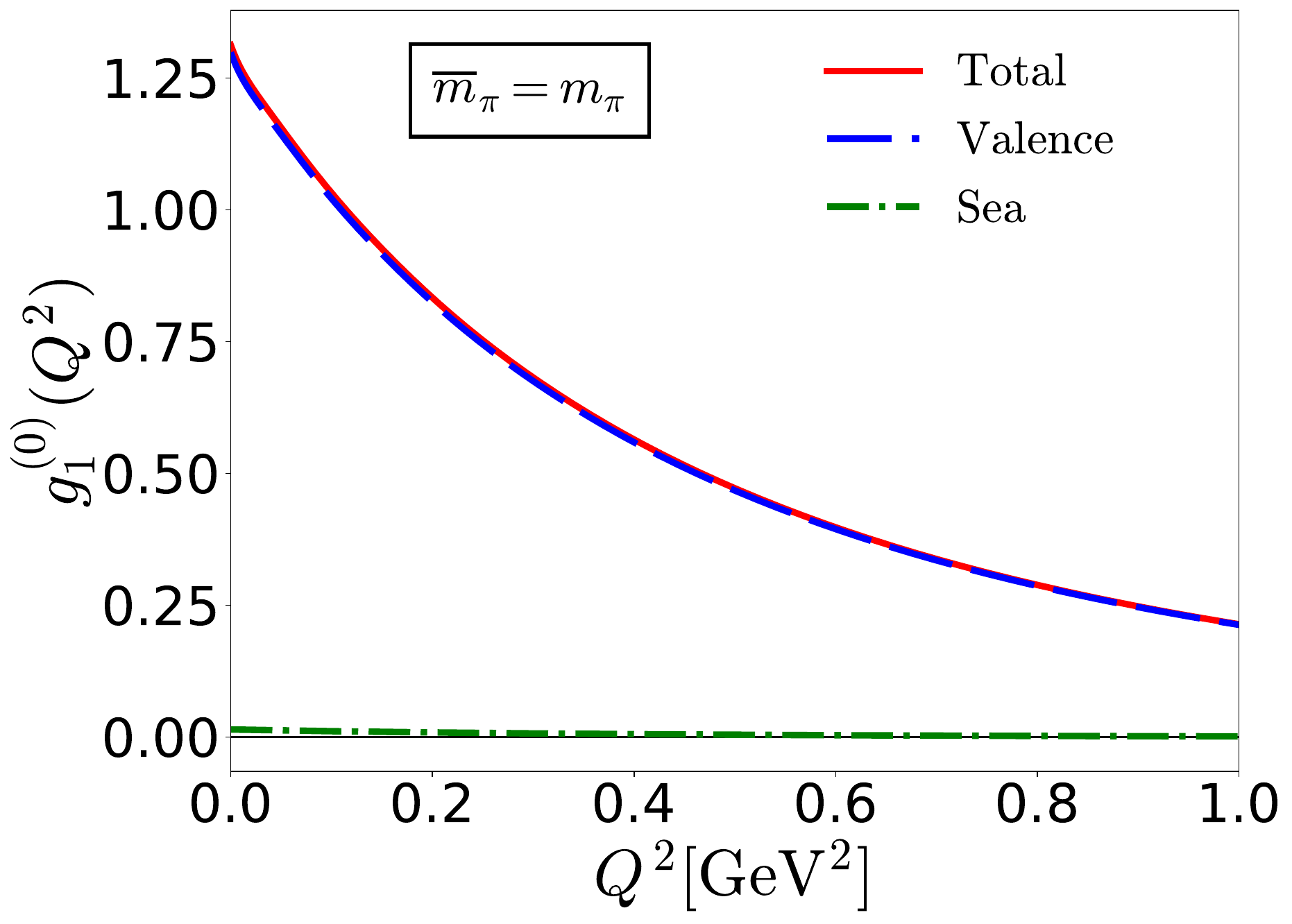}
\includegraphics[scale=0.27]{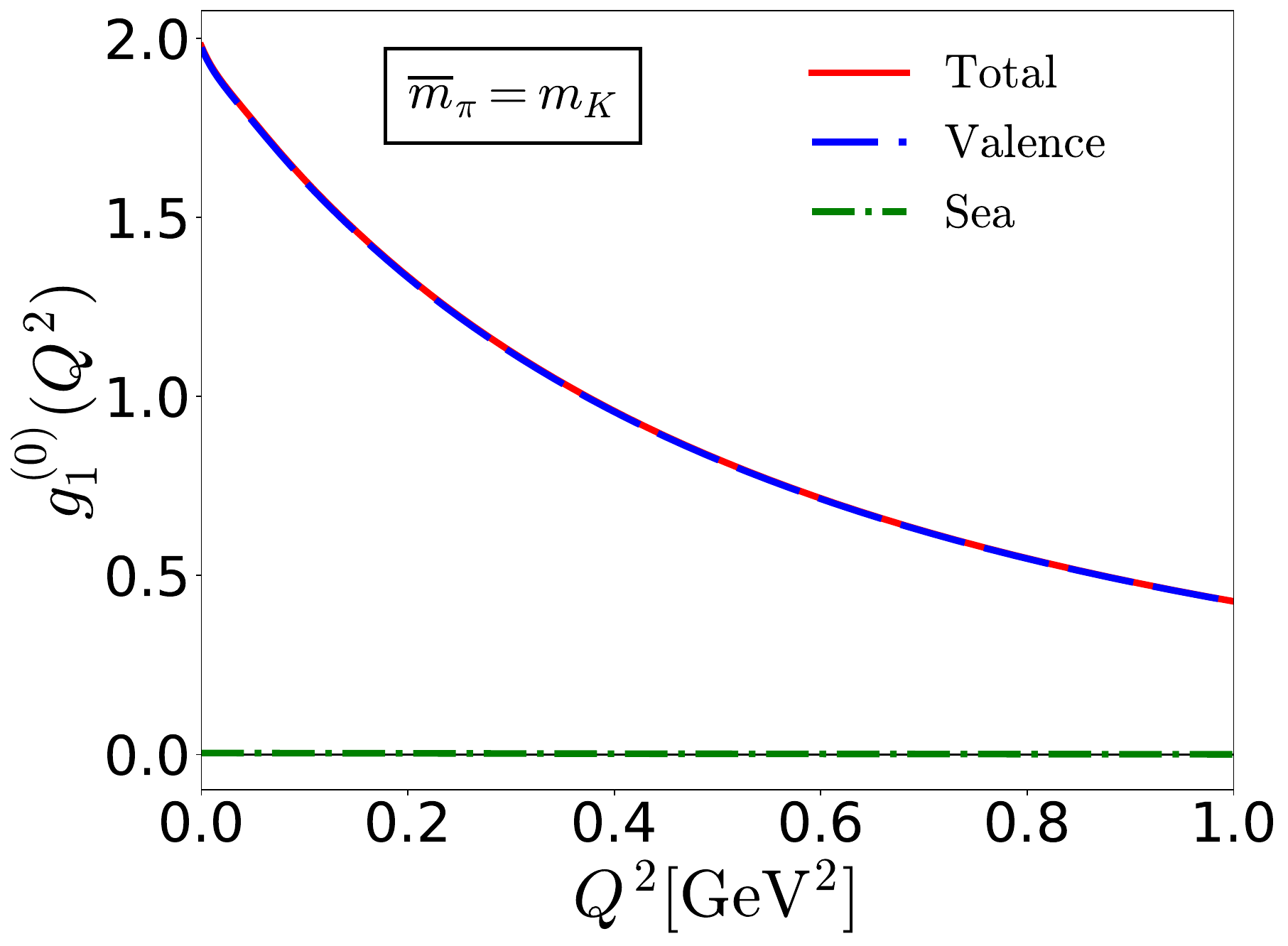}
\includegraphics[scale=0.27]{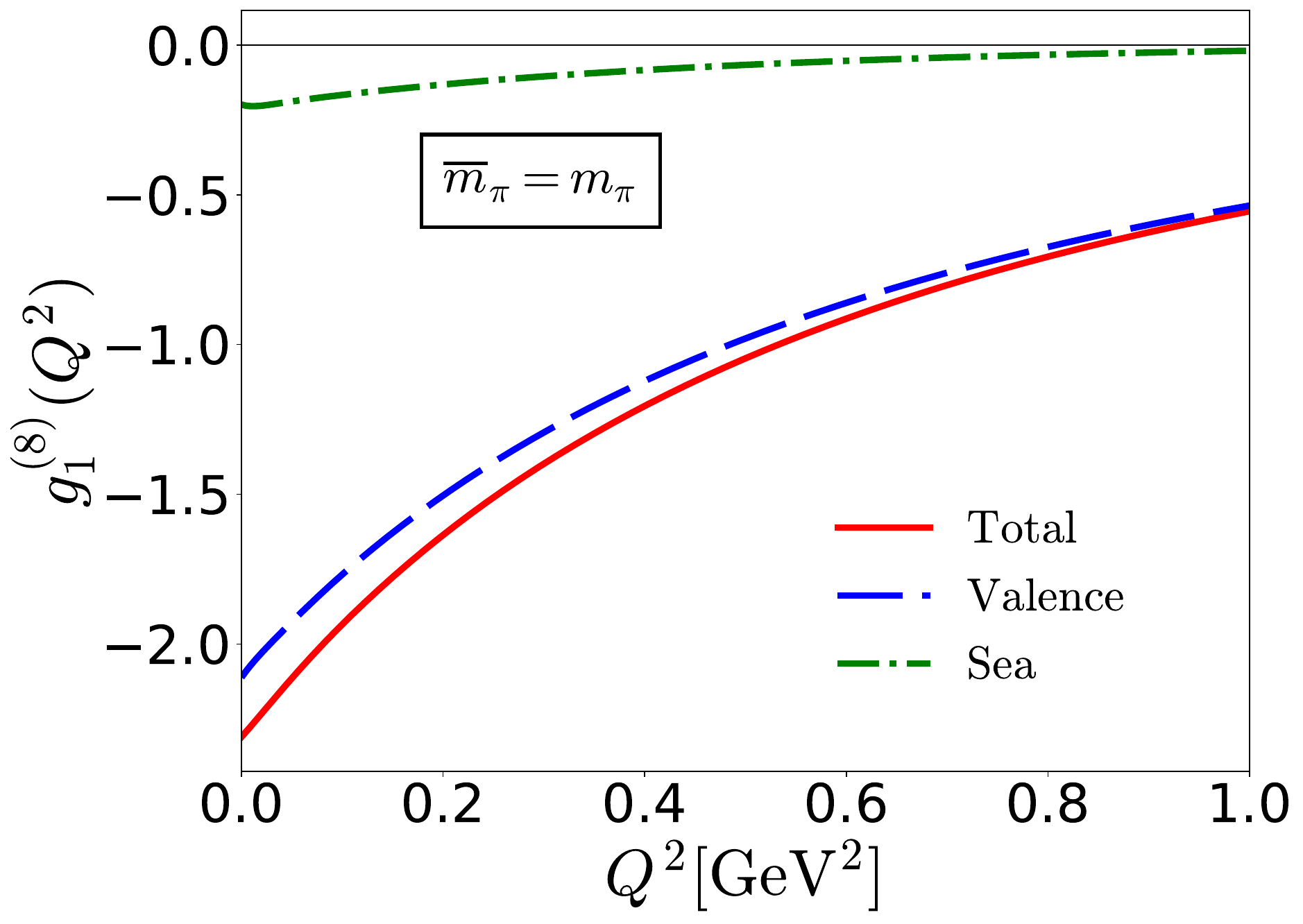}
\includegraphics[scale=0.27]{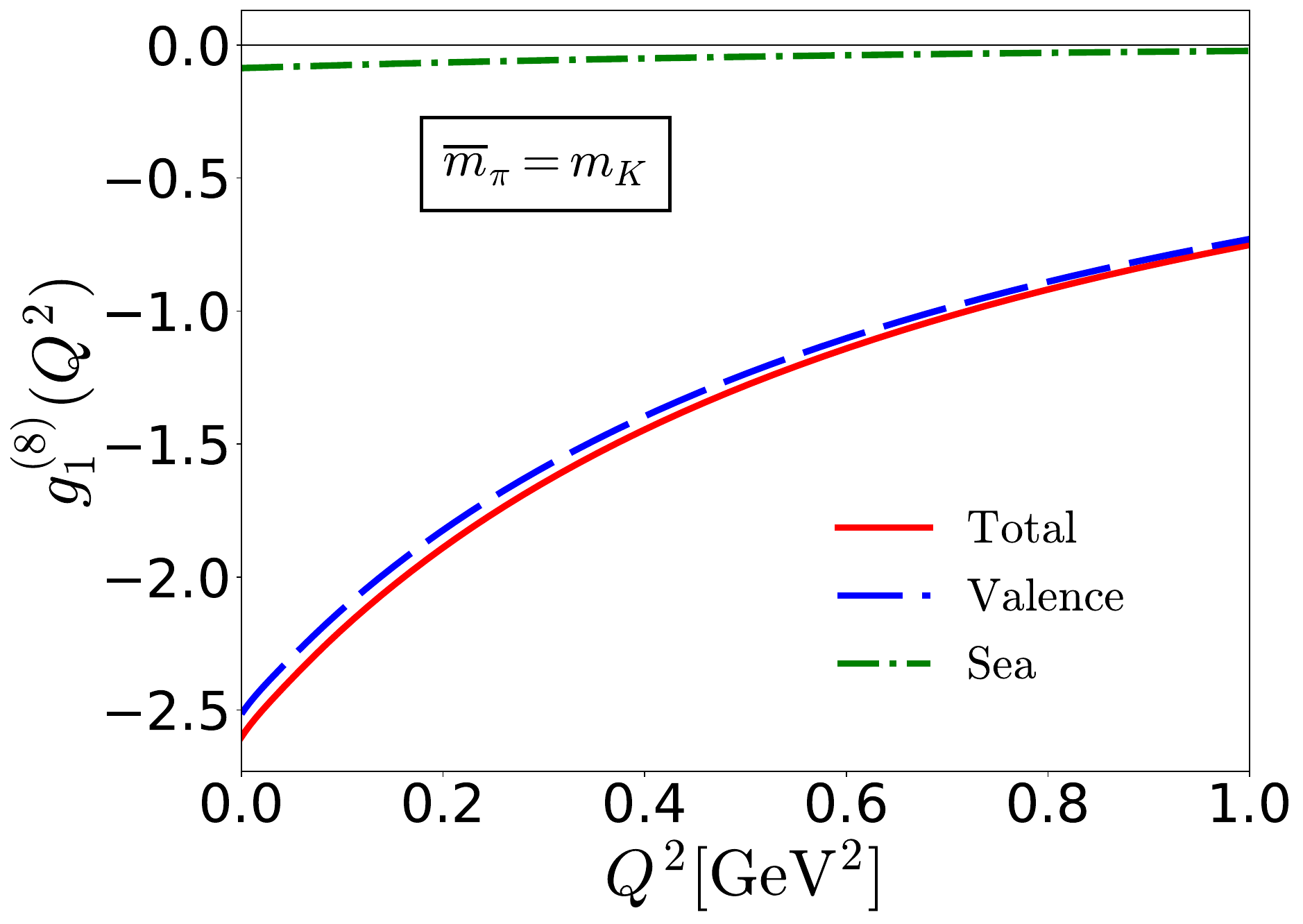}
\caption{Numerical results for the $g_1^{(0)}$ and $g_1^{(8)}$
  axial-vector form factors of the $\Omega^-$ baryon with both the
  pion and kaon clouds considered. In the left panels those with the
  pion cloud are drawn whereas in the right panels those with the kaon
  cloud are depicted. The dashed and dot-dashed curves exhibit the
  valence-quark (level-quark) and sea-quark (Dirac continuum)
  contributions, respectively. The solid curves show the total
  contributions.} 
\label{fig:3}
\end{figure}
\begin{figure}[htp]
\includegraphics[scale=0.27]{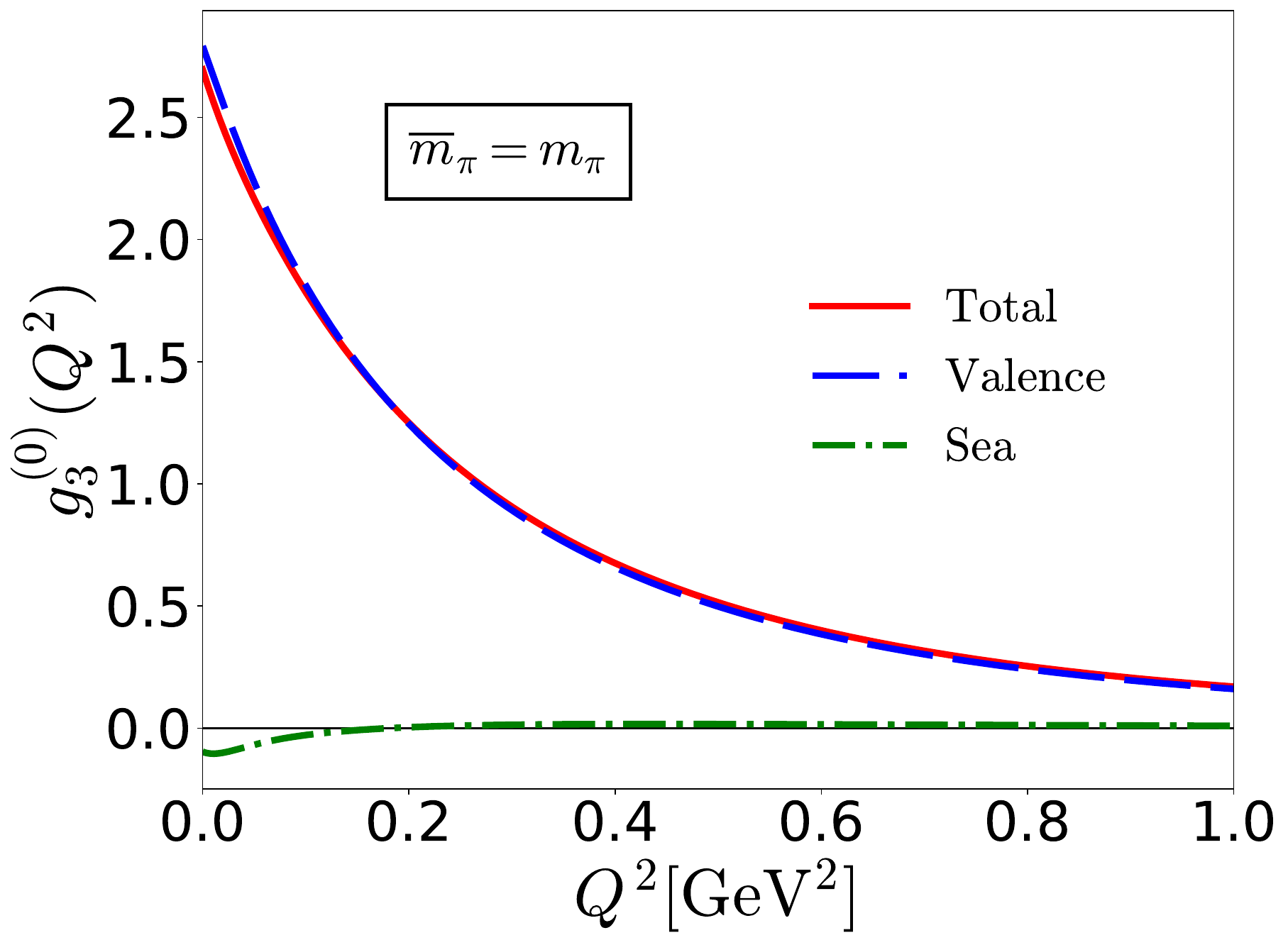}
\includegraphics[scale=0.27]{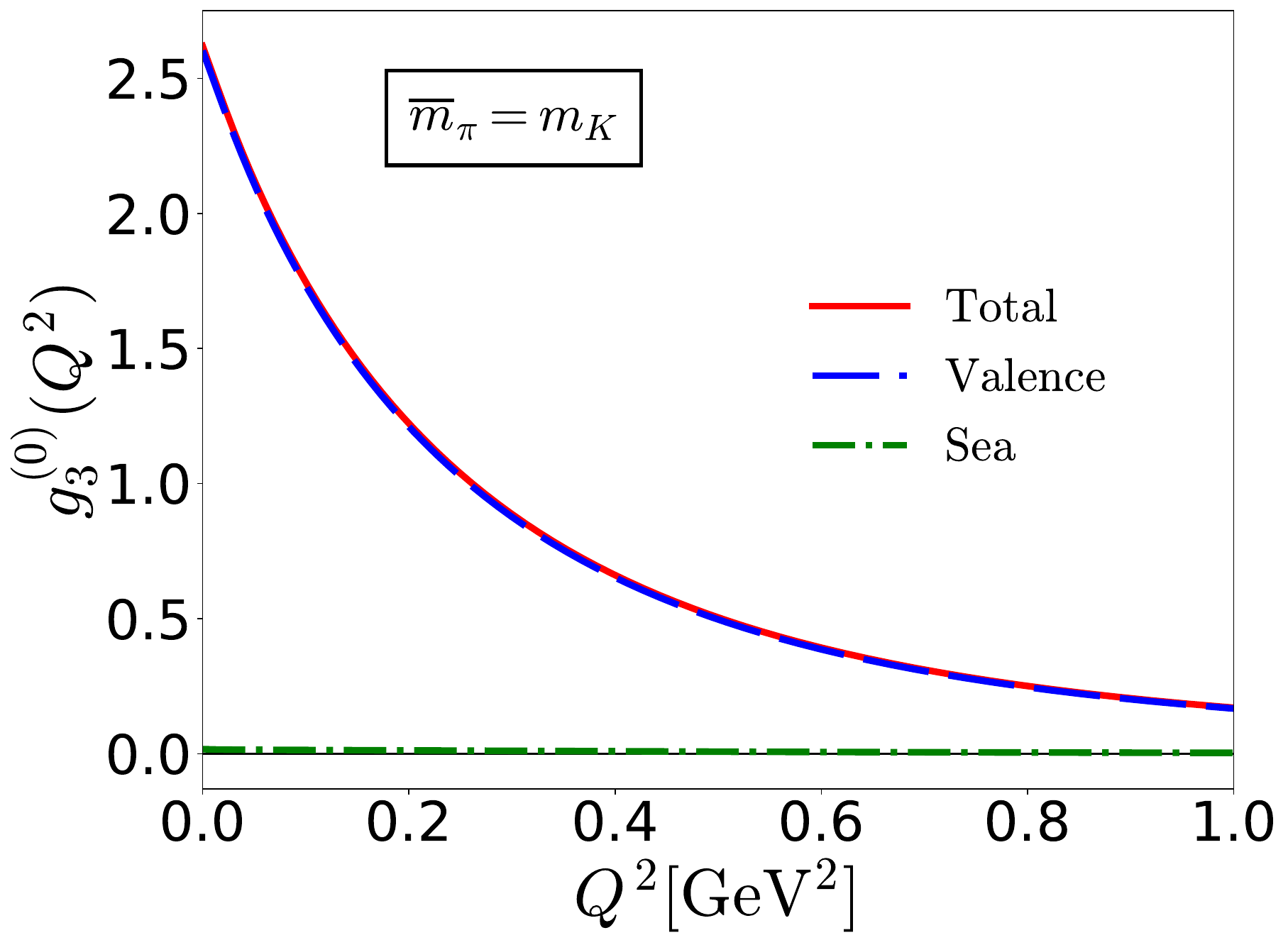}
\includegraphics[scale=0.27]{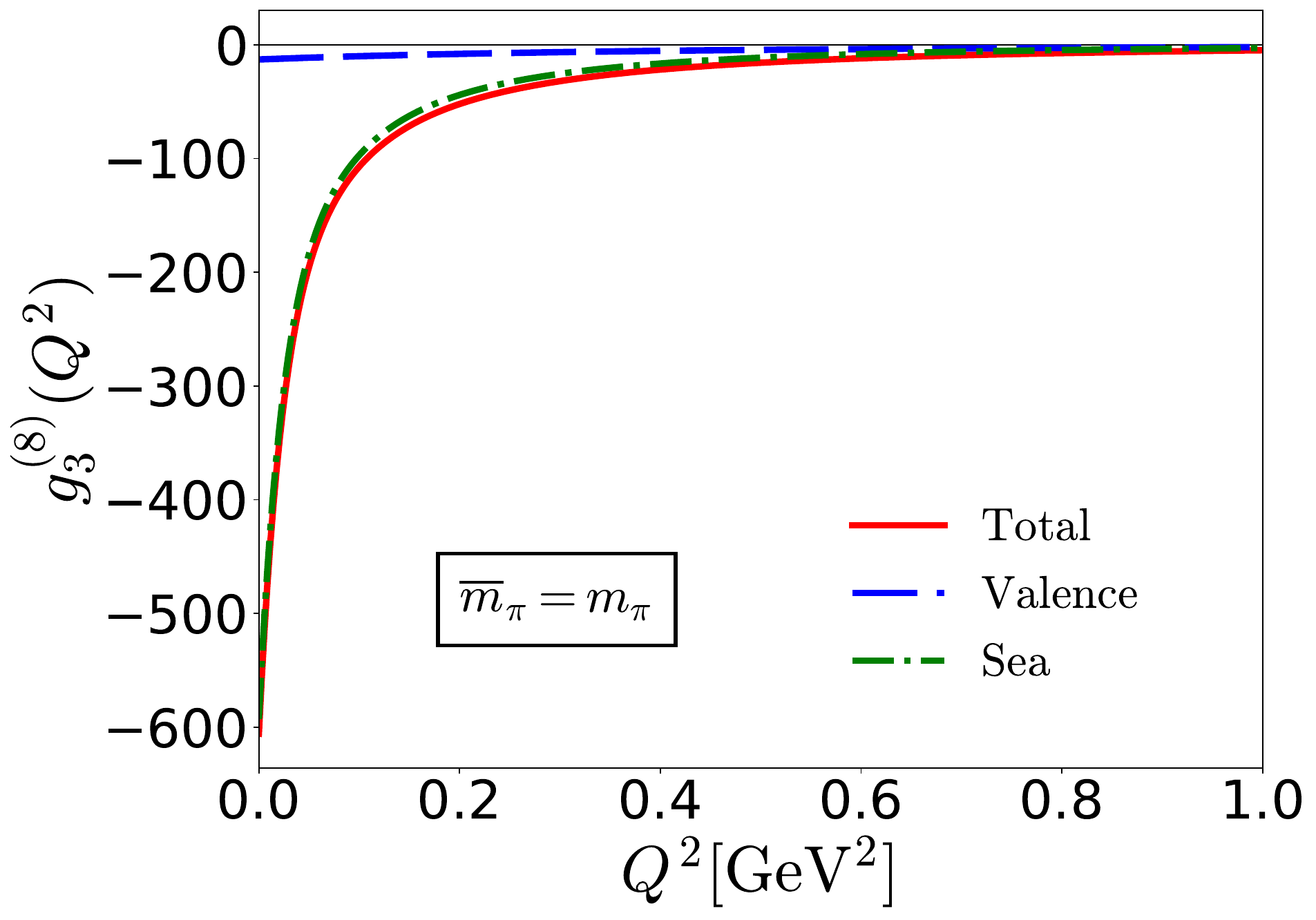}
\includegraphics[scale=0.27]{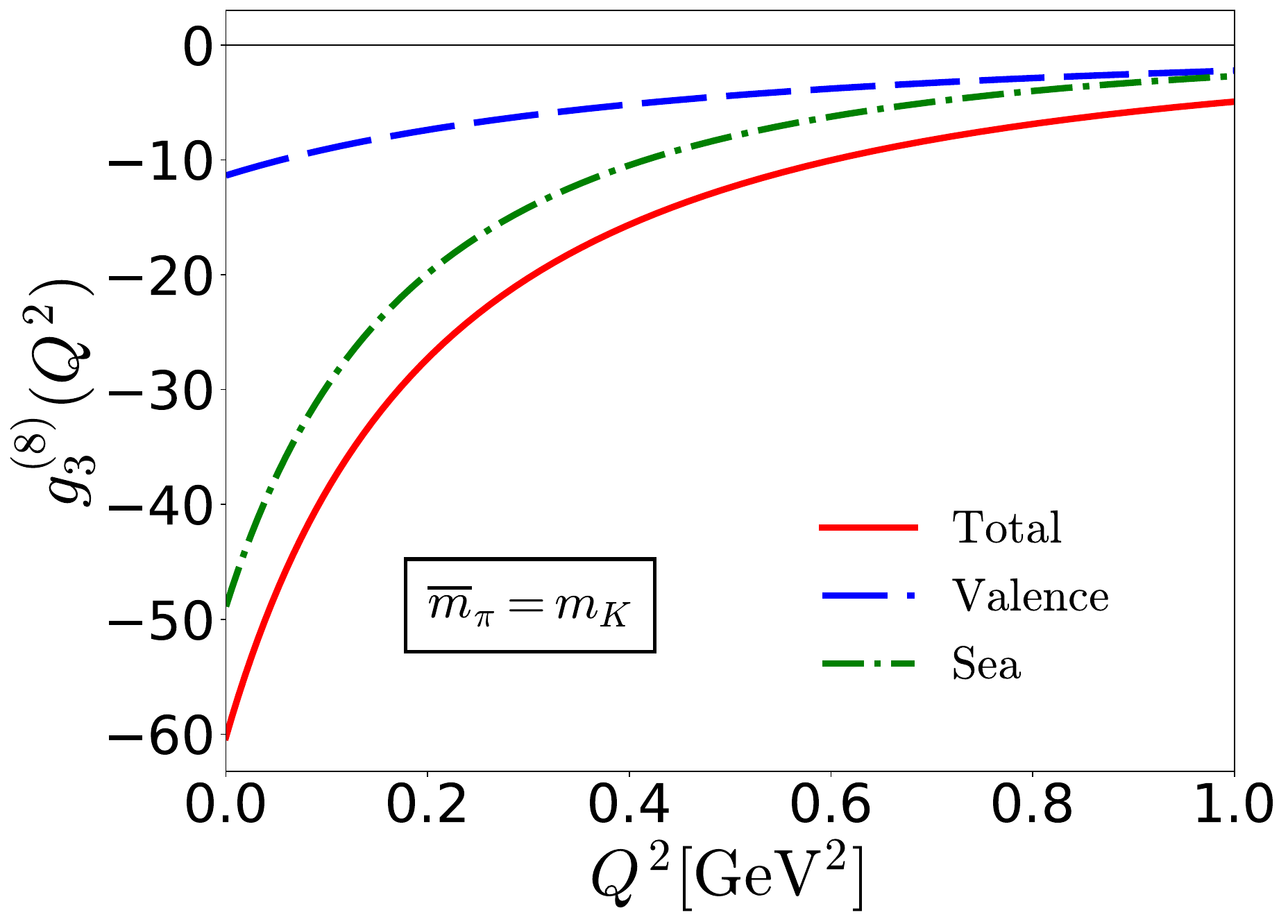}
\caption{Numerical results for the $g_3^{(0)}$ and $g_3^{(8)}$
  axial-vector form factors of the $\Omega^-$ baryon with both the
  pion and kaon clouds considered. In the left panels those with the
  pion cloud are drawn whereas in the right panels those with the kaon
  cloud are depicted. The dashed and dot-dashed curves exhibit the
  valence-quark (level-quark) and sea-quark (Dirac continuum)
  contributions, respectively. The solid curves show the total
  contributions.}
\label{fig:4}
\end{figure}
\begin{figure}[htp]
\includegraphics[scale=0.27]{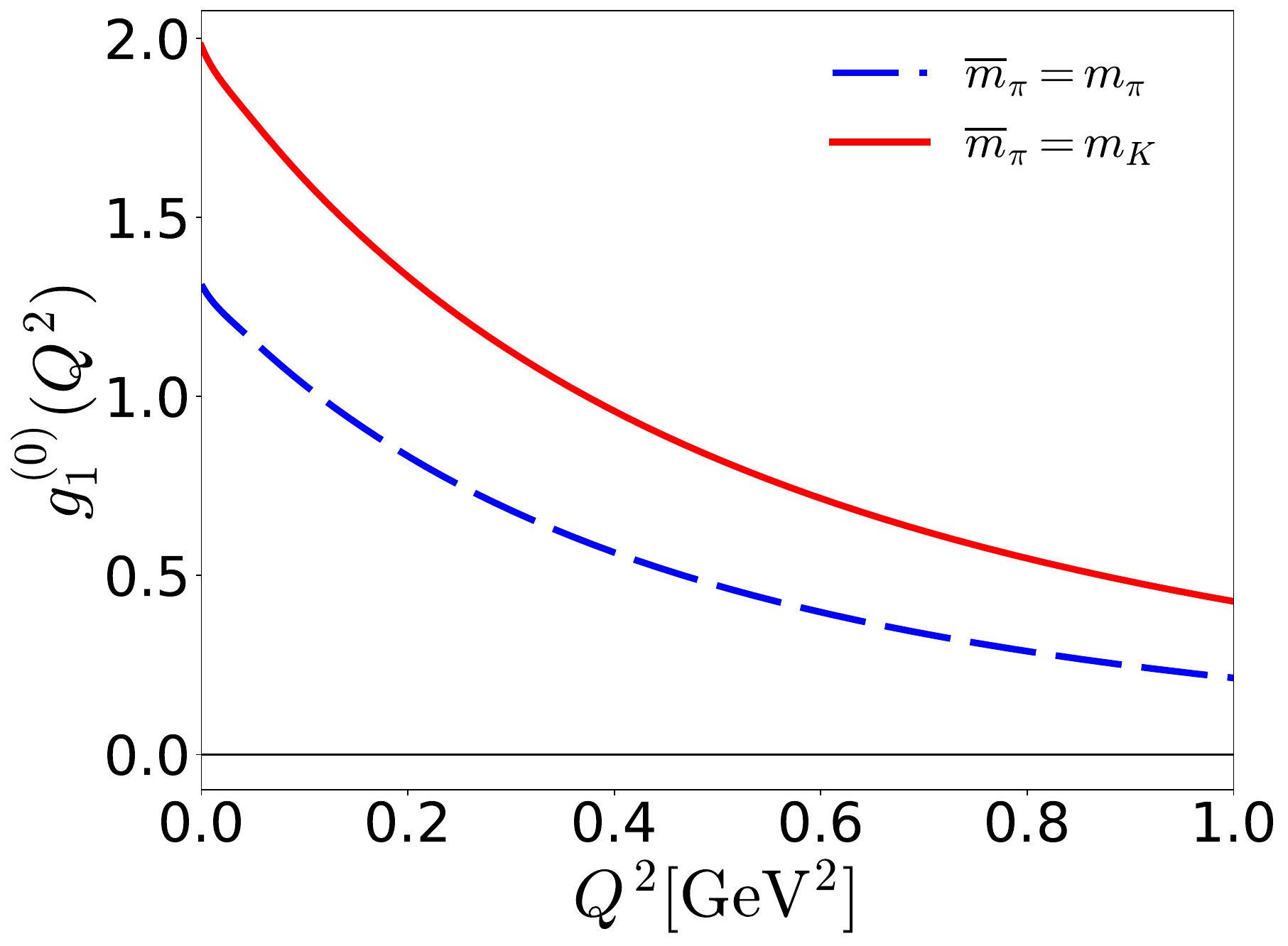}
\includegraphics[scale=0.27]{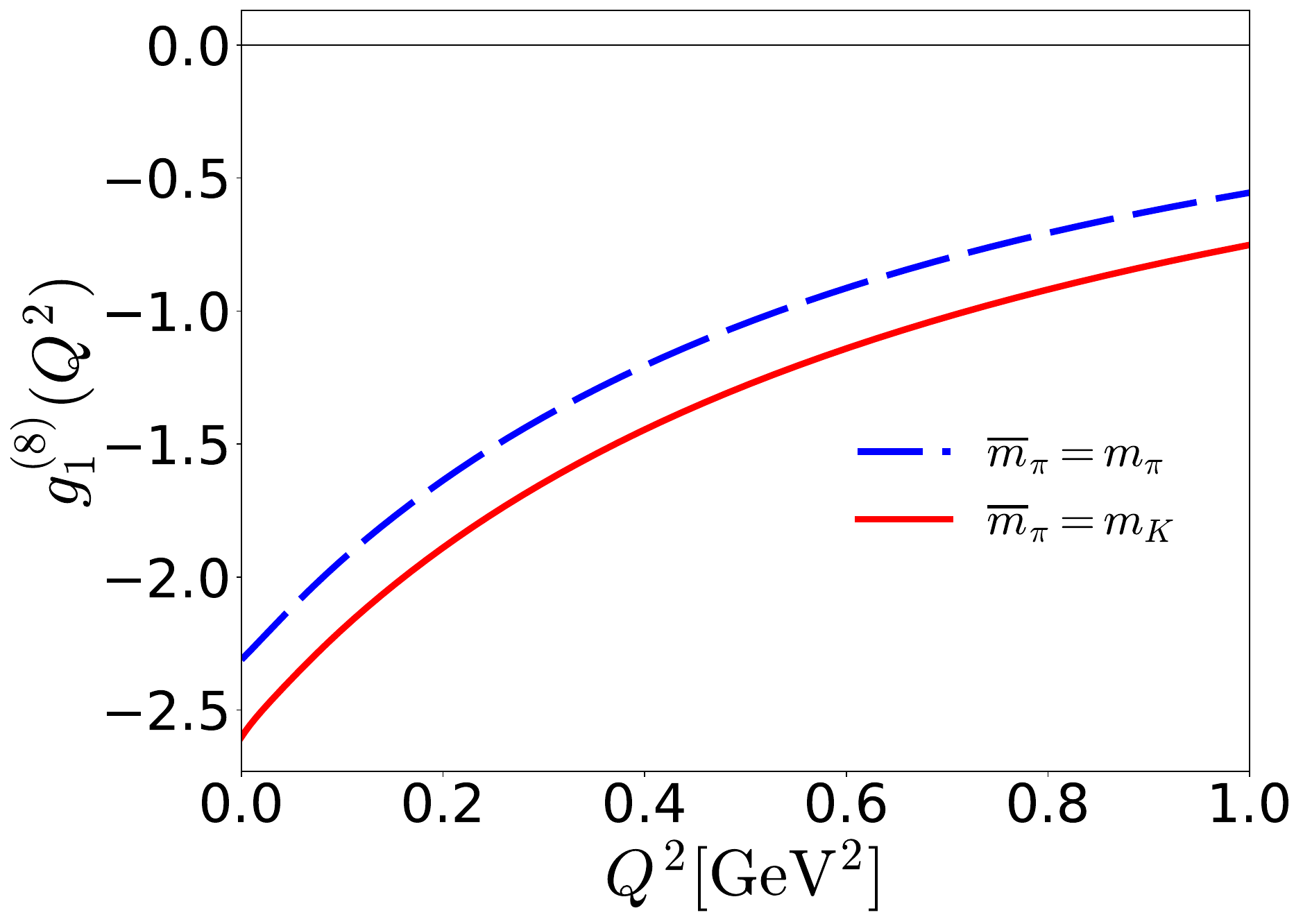}
\includegraphics[scale=0.27]{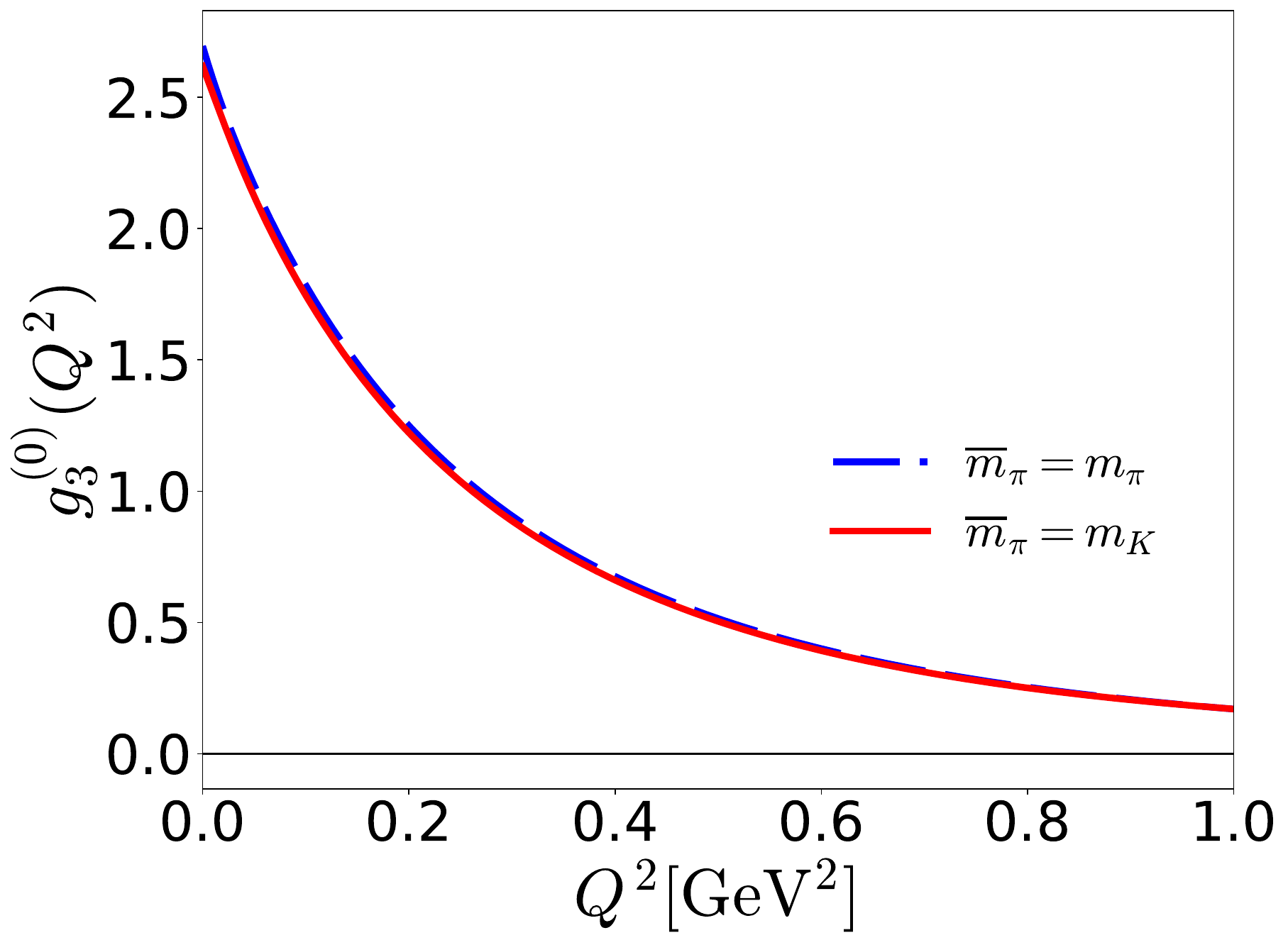}
\includegraphics[scale=0.27]{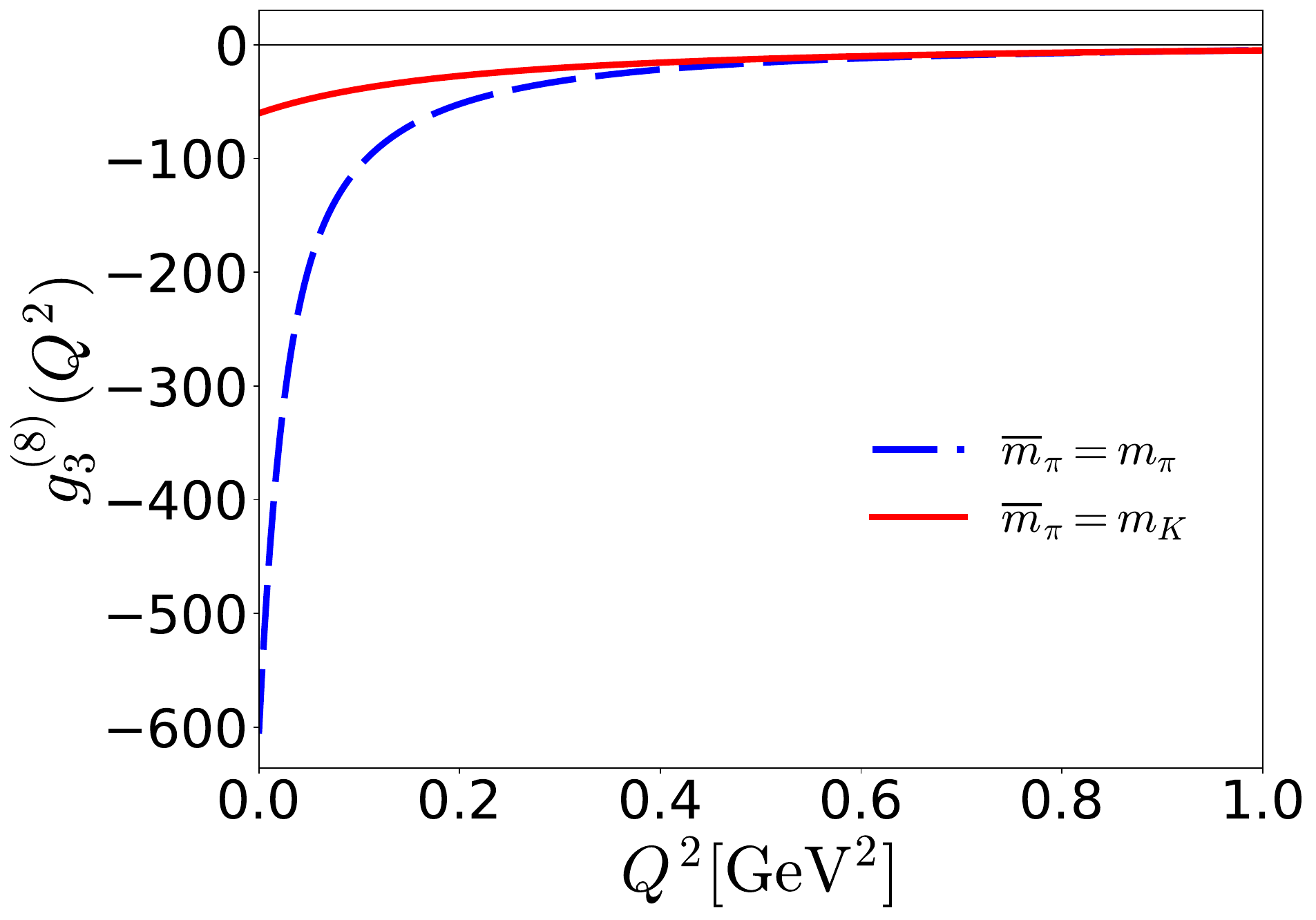}
\caption{Comparison of the present results for the axial-vector form
  factors of the $\Omega^-$. The solid curves 
draw the results with the kaon cloud whereas the dashed ones depict
those with the pion cloud.} 
\label{fig:5}
\end{figure}
We now turn to the axial-vector form factors of the $\Omega^-$
baryon. Since it is an isoscalar baryon, it is not coupled to the 
isovector axial-vector current. So, we first consider
the singlet axial-vector form factors of $\Omega^-$. The axial-vector
form factor $g_1^{(0)}$ is related to the spin content of the $\Omega$
baryon, i.e., its value at $Q^2=0$ has the same physical meaning of
$g_A^{(0)}$ of the nucleon. The value of $g_A^{(0)}$ becomes null in
the Skryme model~\cite{Brodsky:1988ip, Ellis:1988mn}, whereas it
acquires a small value from the rotational $1/N_c$ corrections in the
$\chi$QSM~\cite{Blotz:1993dd, Blotz:1994wi}. The results from the
$\chi$QSM were later improved by employing the symmetry-conserving
quantization~\cite{Praszalowicz:1998jm, Silva:2005fa}. In
Ref.~\cite{Silva:2005fa}, it was shown that the kaon cloud does not
much change the strange component 
of the singlet axial charge and the values of $g_A^{(0)}$, $g_A^{(3)}$
and $g_A^{(8)}$ for the nucleon were obtained respectively to be
$g_A^{(0)}=0.367$, $g_A^{(3)}=1.176$, and $g_A^{(8)}=0.360$.  Subtle
differences between these two models were discussed very in detail in 
Ref.~\cite{Wakamatsu:1996jn}. Figure~\ref{fig:3} illustrates how the
valence- and sea-quark contributions are changed by the replacement of
the pion cloud with the kaon one. Interestingly, the sea-quark
contributions are not much changed by the kaon cloud. However, once
the kaon cloud is considered, the valence-quark contribution makes 
$g_1^{(0)}(Q^2)$ enhanced by about 40~\%. This can be observed clearly
in the upper left panel of Fig.~\ref{fig:5}. On the other hand, the
kaon cloud increases the value of $g_1^{(8)}$ by about 10~\% (see
Figs.~\ref{fig:3} and \ref{fig:5}). It is interesting to compare the
present result for $g_1^{(0)}(Q^2)$ of $\Omega^-$ with that of the
$\Delta^+$ isobar. When the pion cloud is adopted, there is almost no
difference between $g_1^{(0)}(Q^2)$ of the $\Delta^+$ and $\Omega^-$
baryons, as presented in Ref.~\cite{Jun:2020lfx}. However, once we
replace the pion cloud with the kaon one, the situation is drastically
changed. $g_1^{(0)}$ of $\Omega^-$ is about 40~\% larger than that of
$\Delta^+$. This indicates that the up and down sea quarks are less 
likely to pop up in the $\Omega^-$ baryons, compared with the role of
the strange quark in the $\Delta^+$ baryon. So, the sea quarks inside
$\Omega^-$ are less polarized. In the lower panel of 
Fig.~\ref{fig:3}, we depict the results for $g_1^{(8)}$. The
valence-quark contribution is enhanced by about 15~\% whereas that of
the sea-quark is suppressed by about 50~\%, when the kaon cloud was
employed. Altogether, the result for $g_1^{(8)}$ is enhanced by about
12~\% with the kaon cloud used, as shown more clearly in the upper
right panel of Fig.~\ref{fig:5}.  

In the upper panel of Fig.~\ref{fig:4} we draw the numerical results
for $g_3^{(0)}$, which show that $g_3^{(0)}$ is almost not affected by
the replacement of the pion cloud with the kaon one. However, the
kaon cloud reduces remarkably the magnitude of $g_1^{(8)}$ as shown in
the lower panel of Fig.~\ref{fig:4} by about factor 10. In particular,
the contribution of the valence quarks is not much influenced by the
replacement of the pion cloud with the kaon one, whereas the sea-quark
contribution is tremendously lessened. This can be understood by the
fact that the expression for $g_3^{(8)}$ is rather sensitive to the
tail part of the soliton profile function. 
So far, there has been no lattice data on the $g_3^{(8)}$ form factor
of the $\Omega^-$ baryon. It would be very interesting, if one could
compare the present result with the lattice data in the near
future. In Fig.~\ref{fig:5}, we compare the results for the
axial-vector form factors with the kaon cloud to those with the pion
one. Table~\ref{tab:4} summarizes the values of the axial charges of
the $\Omega^-$ baryon, comparing them with the lattice
data~\cite{Alexandrou:2016xok}. With the kaon cloud considered, the
present results for $g_1^{(0)}$ are in better agreement with the
lattice data. However, we do not see any particular improvement for
$g_1^{(8)}$ by replacing the pion cloud with the kaon one. 
\begin{table}[htp]
\caption{Numerical results for the axial charges $g_1^{(0,8)}$
   and $g_3^{(0,8)}$ in comparison with those from lattice QCD
   (LQCD)~\cite{Alexandrou:2016xok}. Since the expressions for the
   octet axial-vector current $A_{\mu}^{8}(x)$ in Ref.~\cite{Alexandrou:2016xok} 
   are different from the present one by $\sqrt{3}$, we have considered 
   it for comparison.}   
\label{tab:4}
\renewcommand{\arraystretch}{1.7}
{\setlength{\tabcolsep}{6pt}
 \begin{tabularx}{0.7\linewidth}{ c | c c c c}
  \hline 
  \hline 
 & $g^{(0)}_{1}(0)$ & $ g^{(8)}_{1}(0)$ 
 & $g^{(0)}_{3}(0)$ & $ g^{(8)}_{3}(0)$ \\
  \hline 
$\overline{m}_{\pi}=m_{\pi}$ & $1.313$ & $-3.863$ & $2.696$ & $-605.734$ \\  
$\overline{m}_{\pi}=m_{K}$ & $1.979$ & $-4.296$ & $2.619$ & $-60.290$ \\  
 \hline
LQCD $(m_{\pi}= 213 \;\mathrm{MeV})$ & $2.0365 \pm 0.0303$ & $-4.0731 \pm 0.0606$ & - & - \\
LQCD $(m_{\pi}= 256 \;\mathrm{MeV})$ & $1.9606 \pm 0.0376$ & $-3.9212 \pm 0.0752$ & - & - \\
LQCD $(m_{\pi}= 302 \;\mathrm{MeV})$ & $2.0215 \pm 0.0441$ & $-4.0431 \pm 0.0883$ & - & - \\
LQCD $(m_{\pi}= 373 \;\mathrm{MeV})$ & $1.9044 \pm 0.0439$ & $-3.8087 \pm 0.0877$ & - & - \\
LQCD $(m_{\pi}= 432 \;\mathrm{MeV})$ & $1.9562 \pm 0.0457$ & $-3.9125 \pm 0.0915$ & - & - \\
 \hline
 \hline
\end{tabularx}}
\end{table}
\section{Summary and conclusions} 
In the present work, we investigated the effects of the kaon cloud on
the electromagnetic and axial-vector properties of the $\Omega^-$
baryon. We first briefly reviewed the mesonic sector, explaining how
the kaonic properties can be described within the present
framework. The cutoff parameters and the average value of the up and
down quark masses were fixed by the pion decay constant and the pion
mass, respectively. Then we were able to reproduce the kaon mass. The 
average quark mass of the up and down quarks inside the one-body Dirac
Hamiltonian produces the asymptotic pion Yukawa tail of the solitonic
profile function and this proper behavior of the Yukawa tail is called
the pion tail. Since the $\Omega^-$ baryon consists of the triple
strange valence quarks, the kaon cloud is required. Thus, we changed
the value of the average quark mass in such a way that the proper kaon
Yukawa tail was produced. While this is theoretically somewhat
inconsistent, it improves phenomenologically the description of the
$\Omega^-$ baryon. As far as we consider only the $\Omega^-$ baryon,
this kaon tail provides even a more plausible theoretical ground than
the pion one. Employing the kaon cloud, we computed the
electromagnetic and axial-vector form factors of the $\Omega^-$
baryon. In the case of the electromagnetic form factors, the
kaon cloud suppresses mainly the sea-quark contributions whereas those
of the valence quarks are not much affected. When the kaon cloud is
considered, the present results for the E0, M1, and E2
form factors are in better agreement with the lattice data in
comparison with those obtained by the pion cloud. The results for the
electric quadrupole moment are also in better agreement with those
from other works. We then calculated the axial-vector form factors of
the $\Omega^-$ baryon and discussed the differences between the
results with the kaon cloud and those with the pion one. The value of
the singlet axial-vector form factor $g_1^{(0)}$ is increased by about
40~\%, whereas the octet one $g_1^{(8)}$ is enhanced by about
12~\%. While the $g_3^{(0)}$ form factor is almost kept intact,
$g_3^{(8)}$ was drastically reduced by about factor 10 when the pion
cloud is replaced with the kaon one.

Since the strange valence quark is deeply related to the kaon cloud,
it is of great importance to examine how the kaon cloud affects all
other hyperons with double strange valence quarks. In particular,
because the singlet axial charges $g_1^{(0)}(0)$ provide information
on the spin content of these  baryons, one has to consider the effects 
of the kaon cloud. The corresponding investigation is under way. 

\begin{acknowledgments}
The authors are very grateful to A. Hosaka for valuable
discussions in the J-PARC Workshop on Physics of Omega Baryons at
the J-PARC-K10 beam line held online for a period of 7-9 June, 2021. The
present work was supported by Basic Science Research Program through
the National Research Foundation of Korea funded by the Ministry of
Education, Science and Technology (Grant-No. 2021R1A2C209336 and
2018R1A5A1025563). J.-Y.K is supported by the Deutscher Akademischer
Austauschdienst(DAAD) doctoral scholarship and in part by BMBF (Grant
No. 05P18PCFP1).  
\end{acknowledgments}

\appendix
\section{ Electromagnetic form factors
  \label{app:a}} 
In Appendix~\ref{app:a}, we compile all the expresstions for the EM
form factors of the $\Omega$ baryon within the framework of the
$\chi$QSM. 
\begin{align}
{G}_{E0}  (Q^{2})= e_{\Omega} \int d^{3} r j_{0}(Q |\bm{r}|)
  \left[\frac{1}{24}{\mathcal{B}}(\bm{r}) + \frac{5}{8I_{1}}
  {\mathcal{I}}_{1}(\bm{r})+\frac{1}{4I_{2}} {\mathcal{I}}_{2}(\bm{r})
  \right],
\label{eq:app1}
\end{align}
where densities of the electric form factor are written explicitly as 
\begin{align}
 \frac{1}{N_{c}}{\cal{B}} (\bm{r})  =&   \langle \mathrm{val}| \bm{r}
 \rangle  \langle \bm{r} | \mathrm{val}
 \rangle -\frac{1}{2}\sum_{n}
 \mathrm{sign}(E_{n})\langle
 n| \bm{r} \rangle \langle \bm{r} | n  \rangle,   \cr 
 \frac{6}{N_{c}}{\cal{I}}_{1} (\bm{r}) =&  \sum_{n \ne
 \mathrm{val}} \frac{1}{E_{n}-E_{\mathrm{val}}}
    \langle \mathrm{val}|
 \bm{r} \rangle \bm{\tau} \langle \bm{r} |
 n  \rangle \cdot \langle n|\bm{\tau}    | \mathrm{val} \rangle +
                                          \frac{1}{2}  \sum_{n ,m}
                                          \langle n| 
   \bm{\tau}  | m \rangle \cdot \langle m| \bm{r} \rangle \bm{\tau}
   \langle \bm{r} | n  \rangle {\mathcal{R}}_{3}(E_{n},E_{m}), \cr 
 \frac{4}{N_{c}} {\cal{I}}_{2} (\bm{r}) =&   \sum_{n^{0}
 }\frac{1}{E_{n^{0}}-E_{\mathrm{val}}}
   \langle n^{0}| \mathrm{val} \rangle \langle \mathrm{val}
 | \bm{r} \rangle  \langle \bm{r} | n^{0}
 \rangle +  \sum_{n ,m^{0}}   \langle n |  m^{0}
   \rangle \langle m^{0}| \bm{r} \rangle 
   \langle \bm{r} | n  \rangle {\mathcal{R}}_{3}(E_{n},E_{m^{0}}).  
\end{align}
The expression for the magnetic dipole form factor of the $\Omega$
baryon is written as   
\begin{align}
{G}_{M1}  (Q^{2})=-\frac{1}{8}e_{\Omega}\frac{ M_{\Omega}}{Q} \int d^{3} r
  \frac{j_{1}(Q |\bm{r}|)}{ |\bm{r}|}
 \left[ \left(  {\cal{Q}}_{0} (\bm{r})  + \frac{1}{I_{1}}
 {\cal{Q}}_{1} (\bm{r}) \right)  +  \frac{1}{2}  \frac{1}{I_{1}}
 {\cal{X}}_{1} (\bm{r}) + \frac{1}{2}
 \frac{1}{I_{2}}  {\cal{X}}_{2} (\bm{r}) \right ],
\label{eq:magfinal}
\end{align}
where the corresponding densities are expressed explicitly as follows: 
\begin{align}
 \frac{1}{N_{c}}{\cal{Q}}_{0} (\bm{r})  =& \langle \mathrm{val}| \bm{r}
 \rangle  \gamma^5\{\bm{r} \times \bm{\sigma} \}\cdot\bm{\tau} \langle
  \bm{r} | \mathrm{val}  \rangle 
  +\sum_{n}   \mathcal{R}_{1}(E_{n}) \langle n| \bm{r} \rangle
  \gamma^5  \{ \bm{r} \times \bm{\sigma} \}\cdot \bm{\tau} \langle
 \bm{r} |  n \rangle,  \cr 
 \frac{2}{N_{c}}{\cal{Q}}_{1} (\bm{r})  =& i  \sum_{n \ne
 \mathrm{val}} \frac{\mathrm{sign}(E_{n})}{E_{n}-E_{\mathrm{val}}}  
 \langle n| \bm{r} \rangle\gamma^5[  \{\bm{r} \times \bm{\sigma}
 \}\times\bm{\tau} ] \langle  \bm{r} | \mathrm{val} \rangle \cdot
 \langle  {\mathrm{val}} | \bm{\tau}| n \rangle
 \cr 
& +i \frac{1}{2}\sum_{n,m}  {\mathcal R}_{4}(E_{n},E_{m})  \langle
  m | \bm{r} \rangle\gamma^5 [\{\bm{r} \times
  \bm{\sigma}\}\times\bm{\tau}] \langle \bm{r} | n \rangle \cdot
  \langle m |\bm{\tau}| n \rangle ,  \cr  
 \frac{1}{N_{c}}{\mathcal{X}}_{1} (\bm{r}) =&  \sum_{n \ne
  \mathrm{val}}\frac{1}{E_{n}-E_{\mathrm{val}}}
   \langle \mathrm{val}| \bm{r} \rangle\gamma^5
  \{\bm{r}\times\bm{\sigma}\} \langle \bm{r} | \mathrm{val}
 \rangle\cdot \langle n|  \bm{\tau} | \mathrm{val}  \rangle \cr  
&+\frac{1}{2}\sum_{n, m} \mathcal{R}_{5}(E_{n},E_{m})\langle n |  
  \bm{r} \rangle \gamma^5  \{\bm{r}\times\bm{\sigma}\}  \langle \bm{r}
  | m  \rangle \cdot  \langle m | \bm{\tau} | n \rangle , \cr 
\frac{1}{N_{c}}{\cal{X}}_{2} (\bm{r})   =&     \sum_{n^{0}}
  \frac{1}{E_{n^{0}}-E_{\mathrm{val}}}
  \langle \mathrm{val}| \bm{r} \rangle\gamma^5 \{\bm{r} \times
  \bm{\sigma} \}\cdot\bm{\tau}  \langle \bm{r} | n^0 \rangle \langle
 n^0   |  \mathrm{val} \rangle  \cr 
& +\sum_{n^0,m}\mathcal{R}_{5}(E_{m},E_{n^{0}}) \langle m |
  \bm{r} \rangle \gamma^5  \{\bm{r} \times
  \bm{\sigma} \}\cdot\bm{\tau} \langle  \bm{r}| n^0 \rangle \langle
  n^0 | m \rangle . 
\label{eq:M1den}                                                   
\end{align}

The expression for the electric quadrupole form factor of $\Omega$
baryon is obtained as  
\begin{align}
G_{E{2}}(Q^{2}) &=  3 \sqrt{{5} } e_{\Omega} M^{2}_{\Omega} \int d^{3} r
                      \frac{j_{2}(Q|\bm{r}|)}{Q^{2}}
                      \left[\frac{1}{I_{2}}\mathcal{I}_{E2}(\bm{r})\right] ,  
\label{eq:magfinal2}
\end{align}
where the densities of the electric quadrupole form factors are
given as 
\begin{align}
(-\sqrt{10})\frac{2}{N_{c}} \mathcal{I}_{1E2}(\bm{r}) =& \sum_{n \ne
\mathrm{val} }\frac{1}{E_{n}-E_{\mathrm{val}}}{\langle\mathrm{val}
| \bm{\tau} | n\rangle} \cdot{\langle n |\bm{r} \rangle
\{ \sqrt{4\pi}Y_{2}  \otimes\tau_{1}  \}_{1}\langle \bm{r} |
\mathrm{val}\rangle} \cr 
&  + \frac{1}{2} \sum_{n,m} \mathcal{R}_{3}(E_n,E_m)   {\langle n |
  \bm{\tau} | m \rangle} \cdot{\langle m | \bm{r} \rangle \{
  \sqrt{4\pi} Y_{2}  \otimes \tau_{1}  \}_{1} \langle \bm{r} | n
  \rangle} ,
\label{eq:E2den}                               
\end{align}
where $|\mathrm{val}\rangle$ and $|n\rangle$ denote the states of the 
valence and sea quarks with the corresponding eigenenergies
$E_{\mathrm{val}}$ and $E_n$ of the single-quark Hamiltonian $h(U_c)$,
respectively~\cite{Christov:1995vm}.
The regularization functions $\mathcal{R}_1(E_n, E_m), \cdots,
\mathcal{R}_5(E_n,E_m)$ will be given at the end of
Appendix~\ref{app:b}.

\section{Axial-vector form factors~\label{app:b}}
The axial-vector form factors of $\Omega^{-}$ in the $\chi$QSM are
expressed as 
\begin{align}
g^{(0)}_{1} (Q^{2}) =& \frac{2N_{c} M_{\Omega}}{9E_{\Omega}} 
  \int d^{3} r \left[ j_{0}(Q |\bm{r}|)\frac{\mathcal{B}_{0}(\bm{r})}{I_{1}}
  -j_{2}(Q |\bm{r}|)\frac{\mathcal{B}_{2}(\bm{r})}{I_{1}} \right], 
\cr
g^{(0)}_{3} (Q^{2}) =& -\frac{8 N_{c} M^{2}_{\Omega}}{9 E_{\Omega} Q^{2}}
  \int d^{3} r 
  \left[j_{0}(Q |\bm{r}|) \left( E_{\Omega}-M_{\Omega} \right) 
  \frac{\mathcal{B}_{0}(\bm{r})}{I_{1}} 
  +j_{2}(Q |\bm{r}|) \left( 2E_{\Omega}+M_{\Omega} \right) 
  \frac{\mathcal{B}_{2}(\bm{r})}{I_{1}} \right], \cr
g^{(8)}_{1} (Q^{2}) =& \frac{\sqrt{3} N_{c} M_{\Omega}}{24E_{\Omega}} 
  \int d^{3} r \left[ j_{0}(Q |\bm{r}|) \left\{ 2\mathcal{A}_{0}(\bm{r})
  -\frac{\mathcal{B}_{0}(\bm{r})}{I_{1}} -\frac{\mathcal{C}_{0}(\bm{r})}{I_{2}} 
  -\frac{i\mathcal{D}_{0}(\bm{r})}{I_{1}} \right\} \right. \cr 
& \hspace{2.6cm} \left. -j_{2}(Q |\bm{r}|) \left\{ 2\mathcal{A}_{2}(\bm{r})
  -\frac{\mathcal{B}_{2}(\bm{r})}{I_{1}} -\frac{\mathcal{C}_{2}(\bm{r})}{I_{2}} 
  -\frac{i\mathcal{D}_{2}(\bm{r})}{I_{1}} \right\} \right],
\cr
g^{(8)}_{3} (Q^{2}) =& -\frac{\sqrt{3} N_{c} M^{2}_{\Omega}}{6 E_{\Omega} Q^{2}}
  \int d^{3} r \left[ j_{0}(Q |\bm{r}|) \left( E_{\Omega}-M_{\Omega} \right) 
  \left\{ 2\mathcal{A}_{0}(\bm{r}) -\frac{\mathcal{B}_{0}(\bm{r})}{I_{1}} 
  -\frac{\mathcal{C}_{0}(\bm{r})}{I_{2}} -\frac{i\mathcal{D}_{0}(\bm{r})}{I_{1}} 
  \right\} \right. \cr 
& \hspace{3.0cm} \left. +j_{2}(Q |\bm{r}|) \left( 2E_{\Omega}+M_{\Omega} \right) 
  \left\{ 2\mathcal{A}_{2}(\bm{r})
  -\frac{\mathcal{B}_{2}(\bm{r})}{I_{1}} -\frac{\mathcal{C}_{2}(\bm{r})}{I_{2}} 
  -\frac{i\mathcal{D}_{2}(\bm{r})}{I_{1}} \right\} \right],
\label{eq:ga}
\end{align}
where 
the components $\mathcal{A}_{0}(\bm{r})$, $\cdots$,
$\mathcal{D}_{0}(\bm{r})$ are written as 
\begin{align}
\mathcal{A}_{0}(\bm{r}) &= \langle \mathrm{val} | \bm{r} \rangle 
  \bm{\sigma} \cdot \bm{\tau} \langle \bm{r} | \mathrm{val} \rangle
  + \sum_{n} \langle n | \bm{r} \rangle \bm{\sigma} \cdot \bm{\tau} 
  \langle \bm{r} | n \rangle \mathcal{R}_{1}(E_n) ,\cr
\mathcal{B}_{0}(\bm{r}) &= \sum_{n \ne \mathrm{val} } 
  \frac{1}{E_{\mathrm{val}}-E_{n}} \langle \mathrm{val} | \bm{r} \rangle \bm{\sigma} 
  \langle \bm{r} | n \rangle \cdot \langle n | \bm{\tau} | \mathrm{val} \rangle 
  -\frac{1}{2} \sum_{n,m} \langle n | \bm{r} \rangle \bm{\sigma} 
  \langle \bm{r} | m \rangle \cdot \langle m | \bm{\tau} | n \rangle 
  \mathcal{R}_{5}(E_n,E_m),\cr 
\mathcal{C}_{0}(\bm{r}) &= \sum_{n_{0} \ne
  \mathrm{val} } \frac{1}{E_{\mathrm{val}}-E_{n_{0}}} 
  \langle \mathrm{val} | \bm{r} \rangle \bm{\sigma} \cdot \bm{\tau} 
  \langle \bm{r} | n_{0} \rangle \langle n_{0} | \mathrm{val} \rangle
  -\sum_{n,m_{0}} \langle n | \bm{r} \rangle \bm{\sigma} \cdot \bm{\tau} 
  \langle \bm{r} | m_{0} \rangle \langle m_{0} | n \rangle 
  \mathcal{R}_{5}(E_n,E_{m_{0}}), \cr
\mathcal{D}_{0}(\bm{r}) &= \sum_{n \ne
  \mathrm{val} } \frac{\mathrm{sgn}(E_{n})}{E_{\mathrm{val}}-E_{n}} 
  \langle \mathrm{val} | \bm{r} \rangle (\bm{\sigma} \times \bm{\tau})
  \langle \bm{r} | n \rangle \cdot \langle n | \bm{\tau} |
                          \mathrm{val} \rangle   
  + \frac{1}{2} \sum_{n,m} \langle n | \bm{r} \rangle \bm{\sigma} 
  \times \bm{\tau} \langle \bm{r} | m \rangle \cdot \langle m | \bm{\tau} 
  | n \rangle \mathcal{R}_{4}(E_n,E_m).
\label{AxComp10}
\end{align}
$\mathcal{A}_{2}(\bm{r})$, $\cdots$, $\mathcal{D}_{2}(\bm{r})$ in
Eq.~\eqref{eq:ga} are given by
\begin{align}
\mathcal{A}_{2}(\bm{r}) =& \langle \mathrm{val} | \bm{r} \rangle 
  \left\{ \sqrt{2\pi}Y_{2} \otimes \sigma_{1}\right\}_{1} \cdot 
  \bm{\tau} \langle \bm{r} | \mathrm{val} \rangle 
  +\sum_{n} \langle n | \bm{r} \rangle \left\{ \sqrt{2\pi}Y_{2} 
  \otimes \sigma_{1}\right\}_{1} \cdot \bm{\tau}
  \langle \bm{r} | n \rangle \mathcal{R}_{1}(E_n) ,\cr
\mathcal{B}_{2}(\bm{r}) =& \sum_{n \ne
  \mathrm{val} } \frac{1}{E_{\mathrm{val}}-E_{n}} 
  \langle \mathrm{val} | \bm{r} \rangle \left\{ \sqrt{2\pi}Y_{2} \otimes 
  \sigma_{1}\right\}_{1} \langle \bm{r} | n \rangle \cdot 
  \langle n | \bm{\tau} | \mathrm{val} \rangle \cr
& -\frac{1}{2} \sum_{n,m} \langle n | \bm{r} \rangle 
  \left\{ \sqrt{2\pi}Y_{2} \otimes \sigma_{1}\right\}_{1} 
  \langle \bm{r} | m \rangle \cdot \langle m | \bm{\tau} | n \rangle 
  \mathcal{R}_{5}(E_n,E_m),\cr 
\mathcal{C}_{2}(\bm{r}) =& \sum_{n_{0} \ne
  \mathrm{val} } \frac{1}{E_{\mathrm{val}}-E_{n_{0}}} 
  \langle \mathrm{val} | \bm{r} \rangle \left\{ \sqrt{2\pi}Y_{2} \otimes 
  \sigma_{1}\right\}_{1} \cdot \bm{\tau} \langle \bm{r} | n_{0} \rangle 
  \langle n_{0} | \mathrm{val} \rangle  \cr
& -\sum_{n,m_{0}} \langle n | \bm{r} \rangle \left\{ \sqrt{2\pi}Y_{2} 
  \otimes \sigma_{1}\right\}_{1} \cdot \bm{\tau} 
  \langle \bm{r} | m_{0} \rangle \langle m_{0} | n \rangle 
  \mathcal{R}_{5}(E_n,E_{m_{0}}), \cr
\mathcal{D}_{2}(\bm{r}) =& \sum_{n \ne
  \mathrm{val} } \frac{\mathrm{sgn}(E_{n})}{E_{\mathrm{val}}-E_{n}} 
  \langle \mathrm{val} | \bm{r} \rangle \left\{ \sqrt{2\pi}Y_{2} \otimes 
  \sigma_{1}\right\}_{1} \times \bm{\tau} 
  \langle \bm{r} | n \rangle \cdot \langle n | \bm{\tau} | \mathrm{val} 
  \rangle \cr
& + \frac{1}{2} \sum_{n,m} \langle n | \bm{r} \rangle 
  \left\{ \sqrt{2\pi}Y_{2} \otimes \sigma_{1}\right\}_{1} 
  \times \bm{\tau} \langle \bm{r} | m \rangle \cdot \langle m | \bm{\tau} 
  | n \rangle \mathcal{R}_{4}(E_n,E_m).
\label{AxComp12}
\end{align}
The regularization functions are defined by
\begin{align}
&\mathcal{R}_{1}(E_{n}) = -\frac{1}{2 \sqrt{\pi}} E_{n}
  \int^{\infty}_{0} \phi(u) \frac{du}{u} e^{-u E_{n}^{2}}, \cr
&\mathcal{R}_{2}(E_{n},E_{m}) = \frac{1}{2 \sqrt{\pi}} \int^{\infty}_{0}
  \phi(u) \frac{du}{\sqrt{u}} \frac{ E_{m} e^{-u E_{m}^{2}} 
  -E_{n}e^{-uE_{n}^{2}}}{E_{n} - E_{m}}, \cr
&\mathcal{R}_{3}(E_{n},E_{m}) = \frac{1}{2 \sqrt{\pi}} \int^{\infty}_{0}
  \phi(u) \frac{du}{\sqrt{u}} \left[ \frac{ e^{-u E_{m}^{2}}- e^{-u
  E_{n}^{2}}}{u(E^{2}_{n} - E^{2}_{m})} -\frac{E_{m} e^{-u
  E_{m}^{2}}+E_{n} e^{-u E_{n}^{2}}}{E_{n} + E_{m}}  \right ], \cr
&\mathcal{R}_{4}(E_{n},E_{m}) = \frac{1}{2 {\pi}} \int^{\infty}_{0}
  \phi(u) {du} \int^{1}_{0} d\alpha e^{-u E_{n}^{2}(1-\alpha) - u
  E^{2}_{m}\alpha}  \frac{E_{n}(1-\alpha) - \alpha
  E_{m}}{\sqrt{\alpha(1-\alpha)}}, \cr
&\mathcal{R}_{5}(E_{n},E_{m}) =
  \frac{\mathrm{sign}(E_{n})-\mathrm{sign}(E_{m})}{2(E_{n}-E_{m})}.
\end{align}



\begin{thebibliography}{99}
\bibitem{Barnes:1964pd}
  V.~E.~Barnes {\it et al.},
  Phys.\ Rev.\ Lett.\  {\bf 12}, 204 (1964).

\bibitem{Biagi:1985rn}
  S.~F.~Biagi {\it et al.},
  Z.\ Phys.\ C {\bf 31}, 33 (1986).

\bibitem{Aston:1987bb}
  D.~Aston {\it et al.},
  Phys.\ Lett.\ B {\bf 194}, 579 (1987).

\bibitem{Aston:1988yn} 
  D.~Aston {\it et al.},
  Phys.\ Lett.\ B {\bf 215}, 799 (1988).

\bibitem{Yelton:2018mag}
  J.~Yelton {\it et al.} [Belle Collaboration],
  Phys.\ Rev.\ Lett.\  {\bf 121}, 052003 (2018).

\bibitem{Jia:2019eav} 
  S.~Jia {\it et al.} [Belle Collaboration],
  Phys.\ Rev.\ D {\bf 100}, 032006 (2019).

\bibitem{Zyla:2020zbs} 
  P.~A.~Zyla {\it et al.} [Particle Data Group],
  PTEP {\bf 2020}, 083C01 (2020).

\bibitem{Acharya:2020asf}
  S.~Acharya {\it et al.} [ALICE Collaboration],
  Nature {\bf 588}, 232 (2020).

\bibitem{Iritani:2018sra}
  T.~Iritani {\it et al.} [HAL QCD Collaboration],
  Phys.\ Lett.\ B {\bf 792}, 284 (2019).

\bibitem{Alexandrou:2010jv}
  C.~Alexandrou, T.~Korzec, G.~Koutsou, J.~W.~Negele and Y.~Proestos,
  Phys.\ Rev.\ D {\bf 82}, 034504 (2010).

\bibitem{Alexandrou:2016xok}
C.~Alexandrou, K.~Hadjiyiannakou and C.~Kallidonis,
Phys. Rev. D \textbf{94}, no.3, 034502 (2016).

\bibitem{Yennie:1957}
D.R. Yennie, M.M. Levy, D.G. Ravenhall, Rev. Mod .Phys. \textbf{29},
144 (1957).  

\bibitem{Frazer:1959gy}
  W.~R.~Frazer and J.~R.~Fulco,
  Phys.\ Rev.\ Lett.\  {\bf 2}, 365 (1959).

\bibitem{Frazer:1960zza}
  W.~R.~Frazer and J.~R.~Fulco,
  Phys.\ Rev.\  {\bf 117}, 1603 (1960).

\bibitem{Frazer:1960zzb}
  W.~R.~Frazer and J.~R.~Fulco,
  Phys.\ Rev.\  {\bf 117}, 1609 (1960).

\bibitem{CohenTannoudji:1972gd}
  G.~Cohen-Tannoudji, V.~V.~Ilyin and L.~L.~Jenkovszky,
  Lett.\ Nuovo Cim.\  {\bf 5S2}, 957 (1972)
   [Lett.\ Nuovo Cim.\  {\bf 5}, 957 (1972)].

\bibitem{Anselm:1972ir}
  A.~A.~Anselm and V.~N.~Gribov,
  Phys.\ Lett.\  {\bf 40B}, 487 (1972).

\bibitem{Adkins:1983hy}
  G.~S.~Adkins and C.~R.~Nappi,
  Nucl.\ Phys.\ B {\bf 233}, 109 (1984).

\bibitem{Frankfurt:1988nt}
  L.~L.~Frankfurt and M.~I.~Strikman,
  Phys.\ Rept.\  {\bf 160}, 235 (1988).

\bibitem{Hammer:2006mw}
  H.-W.~Hammer,
  Eur.\ Phys.\ J.\ A {\bf 28}, 49 (2006).

\bibitem{Meissner:2007tp}
  U.~G.~Meissner,
  AIP Conf.\ Proc.\  {\bf 904}, 142 (2007).

\bibitem{Jenkovszky:2017efs}
  L.~Jenkovszky, I.~Szanyi and C.~I.~Tan,
  Eur.\ Phys.\ J.\ A {\bf 54}, no.7,  116 (2018).

\bibitem{Witten:1979kh}
  E.~Witten,
  Nucl.\ Phys.\ B {\bf 160}, 57 (1979).

\bibitem{Diakonov:1987ty}
  D.~Diakonov, V.~Y.~Petrov and P.~V.~Pobylitsa,
  Nucl.\ Phys.\ B {\bf 306}, 809 (1988).

\bibitem{Diakonov:1997sj}
  D.~Diakonov,
  In *Peniscola 1997, Advanced school on non-perturbative quantum
  field physics* 1-55 [hep-ph/9802298].

\bibitem{Christov:1995vm}
C.~Christov, A.~Blotz, H.-Ch.~Kim, P.~Pobylitsa, T.~Watabe,
T.~Meissner, E.~Ruiz Arriola and K.~Goeke, 
Prog. Part. Nucl. Phys. \textbf{37}, 91-191 (1996).

\bibitem{Kim:1995mr}
  H.-Ch.~Kim, A.~Blotz, M.~V.~Polyakov and K.~Goeke,
  Phys.\ Rev.\ D {\bf 53}, 4013 (1996).

\bibitem{Witten:1983tw}
  E.~Witten,
  Nucl.\ Phys.\ B {\bf 223}, 422 (1983).

\bibitem{Jain:1984gp}
  S.~Jain and S.~R.~Wadia,
  Nucl.\ Phys.\ B {\bf 258}, 713 (1985).

\bibitem{Watabe:1995ts}
T.~Watabe, H.-Ch.~Kim and K.~Goeke,
[arXiv:hep-ph/9507318 [hep-ph]].

\bibitem{Kim:1996vj}
  H.-Ch.~Kim, T.~Watabe and K.~Goeke,
  Nucl.\ Phys.\ A {\bf 616}, 606 (1997).

\bibitem{Silva:2001st}
  A.~Silva, H.-Ch.~Kim and K.~Goeke,
  Phys.\ Rev.\ D {\bf 65}, 014016 (2002)
   Erratum: [Phys.\ Rev.\ D {\bf 66}, 039902 (2002)].

\bibitem{Silva:2002ej}
  A.~Silva, H.-Ch.~Kim and K.~Goeke,
  Eur.\ Phys.\ J.\ A {\bf 22}, 481 (2004).

\bibitem{Silva:2005qm}
  A.~Silva, H.-Ch.~Kim, D.~Urbano and K.~Goeke,
  Phys.\ Rev.\ D {\bf 74}, 054011 (2006).

\bibitem{Yang:2018idi}
G.~S.~Yang and H.-Ch.~Kim,
Phys. Lett. B \textbf{785}, 434 (2018).
  
\bibitem{Praszalowicz:1998jm}
M.~Praszalowicz, T.~Watabe and K.~Goeke,
Nucl. Phys. A \textbf{647}, 49-71 (1999).
  
\bibitem{Jaminon:1989wp}
  M.~Jaminon, G.~Ripka and P.~Stassart,
  Nucl.\ Phys.\ A {\bf 504}, 733 (1989).

\bibitem{Jaminon:1991vz}
  M.~Jaminon, R.~Mendez Galain, G.~Ripka and P.~Stassart,
  Nucl.\ Phys.\ A {\bf 537}, 418 (1992).

\bibitem{Abbott:1981ke}
L.~F.~Abbott,
Acta Phys. Polon. B \textbf{13}, 33 (1982).

\bibitem{Blotz:1992pw}
A.~Blotz, D.~Diakonov, K.~Goeke, N.~W.~Park, V.~Petrov and P.~V.~Pobylitsa,
Nucl. Phys. A \textbf{555}, 765 (1993).

\bibitem{Callan:1987xt}
C.~G.~Callan, Jr., K.~Hornbostel and I.~R.~Klebanov,
Phys. Lett. B \textbf{202}, 269-275 (1988).
\bibitem{Yabu:1987hm}
H.~Yabu and K.~Ando,
Nucl. Phys. B \textbf{301}, 601 (1988).

\bibitem{Kim:2019gka}
J.-Y.~Kim and H.-Ch.~Kim,
Eur. Phys. J. C \textbf{79}, 570 (2019).

\bibitem{Jun:2020lfx}
Y.~S.~Jun, J.~M.~Suh and H.-Ch.~Kim,
Phys. Rev. D \textbf{102}, 054011 (2020).

\bibitem{Capitani:2015sba}
S.~Capitani, M.~Della Morte, D.~Djukanovic, G.~von Hippel, J.~Hua,
B.~J\"ager, B.~Knippschild, H.~B.~Meyer, T.~D.~Rae and H.~Wittig, 
Phys. Rev. D \textbf{92}, 054511 (2015). 

\bibitem{Hansen:2016fzj}
M.~T.~Hansen and S.~R.~Sharpe,
Phys. Rev. D \textbf{93}, 096006 (2016)
[erratum: Phys. Rev. D \textbf{96}, 039901 (2017)].

\bibitem{Alexandrou:2017ypw}
C.~Alexandrou, M.~Constantinou, K.~Hadjiyiannakou, K.~Jansen,
C.~Kallidonis, G.~Koutsou and A.~Vaquero Aviles-Casco, 
Phys. Rev. D \textbf{96}, 034503 (2017).

\bibitem{Kim:2018xlc}
J.~Y.~Kim, H.-Ch.~Kim and G.~S.~Yang,
Phys. Rev. D \textbf{98}, no.5, 054004 (2018).

\bibitem{Kim:2019rcx}
J.~Y.~Kim and H.-Ch.~Kim,
PTEP \textbf{2020}, no.4, 043D03 (2020).



\bibitem{Leinweber:1992hy} 
  D.~B.~Leinweber, T.~Draper and R.~M.~Woloshyn,
  Phys.\ Rev.\ D {\bf 46}, 3067 (1992).



\bibitem{Boinepalli:2009sq} 
S.~Boinepalli, D.~B.~Leinweber, P.~J.~Moran, A.~G.~Williams,
J.~M.~Zanotti and J.~B.~Zhang, 
  Phys.\ Rev.\ D {\bf 80}, 054505 (2009).

\bibitem{Aubin:2008qp} 
  C.~Aubin, K.~Orginos, V.~Pascalutsa and M.~Vanderhaeghen,
  Phys.\ Rev.\ D {\bf 79}, 051502 (2009).



\bibitem{Schlumpf:1993rm} 
  F.~Schlumpf,
  Phys.\ Rev.\ D {\bf 48}, 4478 (1993).

\bibitem{Butler:1993ej} 
  M.~N.~Butler, M.~J.~Savage and R.~P.~Springer,
  Phys.\ Rev.\ D {\bf 49}, 3459 (1994).

\bibitem{Luty:1994ub} 
  M.~A.~Luty, J.~March-Russell and M.~J.~White,
  Phys.\ Rev.\ D {\bf 51}, 2332 (1995).

\bibitem{Lee:1997jk} 
  F.~X.~Lee,
  Phys.\ Rev.\ D {\bf 57}, 1801 (1998).

\bibitem{Wagner:2000ii} 
  G.~Wagner, A.~J.~Buchmann and A.~Faessler,
  J.\ Phys.\ G {\bf 26}, 267 (2000).

\bibitem{Geng:2009ys} 
  L.~S.~Geng, J.~Martin Camalich and M.~J.~Vicente Vacas,
  Phys.\ Rev.\ D {\bf 80}, 034027 (2009).

\bibitem{Li:2016ezv} 
  H.~S.~Li, Z.~W.~Liu, X.~L.~Chen, W.~Z.~Deng and S.~L.~Zhu,
  Phys.\ Rev.\ D {\bf 95}, 076001 (2017).

\bibitem{Krivoruchenko:1991pm} 
  M.~I.~Krivoruchenko and M.~M.~Giannini,
  Phys.\ Rev.\ D {\bf 43}, 3763 (1991).




\bibitem{Oh:1995hn} 
  Y.~s.~Oh,
  Mod.\ Phys.\ Lett.\ A {\bf 10}, 1027 (1995).

\bibitem{Buchmann:2002et} 
  A.~J.~Buchmann and R.~F.~Lebed,
  Phys.\ Rev.\ D {\bf 67}, 016002 (2003).



\bibitem{Azizi:2008tx}
  K.~Azizi,
  Eur.\ Phys.\ J.\ C {\bf 61}, 311 (2009).

\bibitem{Aliev:2009pd} 
  T.~M.~Aliev, K.~Azizi and M.~Savci,
  Phys.\ Lett.\ B {\bf 681}, 240 (2009).

\bibitem{Brodsky:1988ip}
S.~J.~Brodsky, J.~R.~Ellis and M.~Karliner,
Phys. Lett. B \textbf{206}, 309-315 (1988).

\bibitem{Ellis:1988mn}
J.~R.~Ellis and M.~Karliner,
Phys. Lett. B \textbf{213}, 73-80 (1988).

\bibitem{Blotz:1993dd}
A.~Blotz, M.~Praszalowicz and K.~Goeke,
Phys. Lett. B \textbf{317}, 195-199 (1993).

\bibitem{Blotz:1994wi}
A.~Blotz, M.~Praszalowicz and K.~Goeke,
Phys. Rev. D \textbf{53}, 485-503 (1996).

\bibitem{Silva:2005fa}
A.~Silva, H.-Ch.~Kim, D.~Urbano and K.~Goeke,
Phys. Rev. D \textbf{72}, 094011 (2005).

\bibitem{Wakamatsu:1996jn}
M.~Wakamatsu,
Prog. Theor. Phys. \textbf{95}, 143-173 (1996)
\end{thebibliography}
\end{document}